\documentclass[12pt,a4paper,twoside]{article}
\usepackage{graphicx}
\usepackage{amssymb,latexsym,amsfonts}

\newcommand{\be}{\begin{equation}}
\newcommand{\ee}{\end{equation}}
\newcommand{\bd}{\begin{displaymath}}
\newcommand{\ed}{\end{displaymath}}
\newcommand{\ba}{\begin{array}}
\newcommand{\ea}{\end{array}}
\newcommand{\bt}{\begin{tabular}}
\newcommand{\et}{\end{tabular}}

\newcommand{\bea}{\begin{eqnarray}}
\newcommand{\eea}{\end{eqnarray}}
\newcommand{\bean}{\begin{eqnarray*}}
\newcommand{\eean}{\end{eqnarray*}}

\newcommand{\mc}{\multicolumn}

\newcommand{\hlf}{\frac{1}{2}}

\newcommand{\dif}{\mathrm{d}}

\newcommand{\inp}[2]{\langle #1, #2 \rangle}

\newcommand{\Z}{\mathbb{Z}}

\newcommand{\R}{\mathbb{R}}
\newcommand{\C}{\mathbb{C}}

\newcommand{\mod}{\textrm{mod}}

\newcommand{\df}[1]{{} * \! #1}
\newcommand{\tp}[1]{ #1^{+++}}

\begin{document}
\begin{titlepage}

\begin{center}

\vskip 1 cm
{\Large \bf Poincar\'e Duality\\
\Large  \large and\\
\Large $\tp{G}$ algebra's\\}
\vskip 1.25 cm
{Arjan Keurentjes${}^{\dag *}$}\\
\vskip 0.5cm
{\sl Theoretische Natuurkunde, Vrije Universiteit Brussel,\\ 
and \\
The International Solvay institutes,\\
Pleinlaan
  2, B-1050 Brussels, Belgium \\}
\end{center}
\vskip 2 cm
\begin{abstract}
\baselineskip=18pt
Theories with General Relativity as a sub-sector exhibit enhanced symmetries upon dimensional reduction, which is suggestive of ``exotic dualities''. Upon inclusion of time-like directions in the reductions one can dualize to theories in different space-time signatures. We clarify the nature of these dualities and show that they are well captured by the properties of infinite-dimensional symmetry algebra's ($\tp{G}$-algebra's), but only after taking into account that the realization of Poincar\'e duality leads to restrictions on the denominator subalgebra appearing in the non-linear realization. The correct realization of Poincar\'e duality can be encoded in a simple algebraic constraint, that is invariant under the Weyl-group of the $\tp{G}$-algebra, and therefore independent of the detailed realization of the theory under consideration. We also construct other Weyl-invariant quantities that can be used to extract information from the $\tp{G}$ algebra without fixing a level decomposition. 
\end{abstract}

\vskip 1cm

{\small ${}^\dag$ e-mail address: arjan@tena4.vub.ac.be \\
${}^*$ Post-doctoraal onderzoeker van het Fonds voor Wetenschappelijk Onderzoek, Vlaanderen}
\end{titlepage}
\section{Introduction}

One of the significant features of string theory is T-duality: Strings compactified on a large circle exhibit the same spectrum as strings compactified on a small circle. There are several arguments indicating that this is an exact symmetry of the full non-perturbative theory (for reviews and original references see \cite{T}).

A remnant of T-duality is present at the level of the low-energy
effective action, of the reduced theory, where all the massive string modes have been thrown away. It manifests itself as an ambiguity: Given the low-energy effective action, there are multiple, inequivalent ways of interpreting it as a reduction of a higher dimensional theory. Applying these ambiguities to solutions of the reduced theory allows for solution-generating techniques. T-duality (and U-duality) however goes one step further, and claims that some of the solutions related by these "accidental" symmetries actually are equivalent.

The (bosonic sector of the) dimensionally reduced low-energy action describes a theory of General relativity coupled to various form fields. One expects the appearance of symmetries that mix the directions over which one is reducing, but the theory may exhibit enhanced, previously "hidden" symmetries \cite{Cremmer:1979up, Julia:1980gr, Julia:1982gx}. Such enhanced symmetries render the assignment of a subgroup of "geometrical" symmetries ambiguous, and are therefore a clear indication of the existence of dual formulations. They are however not an exclusive feature of the low-energy actions of string theories and M-theories; they occur in a much wider class of theories, of which many seem to have no direct relation with string- or M-theory \cite{Breitenlohner:1987dg, Cremmer:1999du, Keurentjes:2002xc, Keurentjes:2002rc}. Most notable, among these theories are the "bare" theories of General Relativity, as first discovered by Ehlers. A natural question is to what extent one can make sense of these duality symmetries.

A simple observation is that, if a theory of form fields is coupled to gravity, these exhibit $p$-brane solutions describing the gravitational back-reaction to some extended object (see \cite{Ortin:2004ms} for a pedagogical review and an extensive reference list). Compactifying such an extended object on a torus of suitable dimension, one finds an effective theory describing strings, which possibly should exhibit T-dualities. This argument is admittedly very na\"\i ve, as it is well known that the higher dimensional action should also contain suitable couplings to dilatons and Chern-Simons terms if the reduced theory is to be regarded as a string theory. Even if we fine-tune for these, the resulting theory will in general be a theory of which it is unclear whether there exists a consistent, non-perturbative quantum extension. We will however ignore such questions and focus on necessary conditions for the presence of T-dualities from the classical theory. It will be clear then that non-linearly realized, enhanced symmetries and $p$-brane solutions are vital ingredients. This discussion even applies to (higher-dimensional) General Relativity, where there is at least one category of suitable solutions: The extremal Kaluza-Klein monopoles are for theories in 5 or more dimensions extended objects, and can be compactified to 5 dimensions to make string-like objects.

An interesting development was the application of T-duality to time-like circles \cite{Hull:1998vg, Hull:1998ym, Hull:1998fh}. Under such dualities, it is not at all guaranteed that a space-like circle gives rise to a space-like circle in a dual picture, or that a time-like circle will dualize to a time-like circle. In these circumstances the space-time signature is no longer a fixed quantity, but is dependent on the various dual points of view. Again these dual points of view should be reflected in the symmetries of the low-energy effective action \cite{Cremmer:1998em, Hull:1998br, Hull:1998ym, Keurentjes:2004xx}.

For the ultimate unified point of view one would like to have a formulation with a symmetry that unifies all the possible space-like and time-like reductions, and the possible dualities. For the special situation where the reduction to 3-dimensions of the theory exhibits a non-linearly realized symmetry $G$ \cite{Breitenlohner:1987dg, Cremmer:1999du, Keurentjes:2002xc, Keurentjes:2002rc}, it has been proposed that candidate symmetry algebra's should contain a "triple extended" Lie algebra $\tp{G}$ \cite{West:2001as, Gaberdiel:2002db, West:2002jj, Englert:2003zs, Kleinschmidt:2003mf}. The extended algebra $\tp{G}$ is built (with the standard rules \cite{Kac:1990gs}) from its Dynkin diagram, which is basically the Dynkin diagram of $G$ extended with 3 more nodes. According to the conjectures of \cite{West:2002jj, Englert:2003zs, Kleinschmidt:2003mf},  the degrees of freedom of the theory should be described by $\tp{G}/K(\tp{G})$, where $K(\tp{G})$ is generated by a sub-algebra of $\tp{G}$ invariant under a particular involution. The more conventional formulation of the theory should be retrieved by a kind of level-decomposition, where a "gravitational sub-algebra" is singled out.

It was soon realized that the space-time signature is encoded in (a subgroup of) $K(\tp{G})$ \cite{West:2003fc, Englert:2003py, Schnakenburg:2003qw, Keurentjes:2003hc, Keurentjes:2004bv}. It however turns out that the space-time signature in these theories is not fixed, but depends on the choice of gravitational sub-algebra \cite{Keurentjes:2004bv}. Also outside the Lorentz-subalgebra of $K(\tp{G})$ there may be additional signs that should be reflected in the kinetic terms of the form fields \cite{Keurentjes:2004bv}. This is actually the algebraic translation of the time-like T-dualities that have subsequently been studied in this algebraic context \cite{Englert:2004ph, deBuyl:2005it, Cook:2005wj}.

In the present paper we would like to point out that there is one more aspect that should play a role in the analysis. A priori it seems as if $\tp{G}$ algebra's can accommodate for any space-time signature and many possibilities for signs of the gauge-fields \cite{Keurentjes:2004bv}. On second thought, this should be a cause of concern: In $\tp{G}$ algebra's forms and their duals occur independently, though the signs of forms and their duals should not be independent; and theories with (anti-)self dual tensors can only exist in space-times of appropriate signature. Both issues are related to Poincar\'e duality, which is not, but should be implemented somehow on these algebra's. In the present paper we find a simple criterion which as we will prove implements Poincar\'e duality correctly. The same ingredient is needed crucially to ensure the correct transformation properties of the theory under T-duality.

In section \ref{dual} we will collect the relevant facts on Poincar\'e
duality and T-duality. The allowed signature changes actually follow a
very simple pattern that can be analyzed without any reference to
$\tp{G}$ algebra's or string theory. This is independent of, but
should be implemented correctly on the $\tp{G}$-algebra's. Section
\ref{tech} recalls and explains why Weyl-reflections and other diagram
automorphisms correspond to dualities. We will also sketch two, in our
view, important complications in the treatment of $G/H$ sigma models
which restrict explicit computations, but can be avoided in an abstract approach.
In section \ref{Ggeneral} we will briefly review the implementation of space-time signature and other signs in the context of $\tp{G}$-theories. Then we will derive a simple but universal criterion that ensures that the theory has the correct transformation properties under Poincar\'e duality. The criterion is formulated as a constraint on a quantity that is invariant under Weyl reflections and outer automorphisms, and hence in one go restricts all dual formulations of the theory. For specific $\tp{G}$ algebra's there is a larger class of such invariants (though not necessarily invariant under outer automorphisms), for which we give a construction recipe. Section \ref{Gspecific} discusses the specific theories with their invariants. In the concluding section \ref{Con} we will summarize and discuss our results.

\section{Duality from dimensional reduction and oxidation} \label{dual}

The aim of the present section is to give an intuitive feeling what the meaning of T-duality and in particular time-like T-duality is outside the context of string theory. For this we will (ab-)use the low energy effective action. This section will be rather sketchy in nature, but we will not need much detail to derive precise statements about which signature changes should be allowed. 

\subsection{Sources of signs}

The discussion that follows will be essentially one about minus signs. In the context of dimensional reduction and oxidation, such minus signs are essentially contributed by 3 sources
\begin{enumerate}
\item The space-time signature, that we will leave arbitrary unless
  indicated otherwise. This assigns different signs to timelike and
  space-like directions in the Lorentz-invariant inner product. After
  fixing a space-time signature, the Lorentz-invariant inner product
  on the tangent space assigns definite signs to contractions of tensors. Our convention will be to assign to time-like directions a minus sign in the Lorentz-invariant inner product, and to space-like directions a plus sign.
\item Explicit signs. The choice of space-time signature fixes the implicit signs present in the contraction of field strengths for the various gauge fields to a Lorentz-invariant combination, but not the overall sign. Though it is customary to choose a particular sign to avoid "ghosts" (unphysical modes) in quantum theory, we will leave also these arbitrary unless indicated otherwise.
\item Poincar\'e duality. The non-linear symmetries present in dimensionally reduced theories of General Relativity coupled to various form-fields, mix fields and their Poincar\'e duals. It is a simple exercise to derive that (in the absence of Chern-Simons terms and other complicating factors) \cite{Hull:1998br}
\be
\int d^D x \ (F_{(n)})^2 \rightarrow (-)^{T-1} \int d^D x \ (F_{(D-n)})^2. 
\ee
Hence if the number of time directions is even, either the form or its dual carries a kinetic term with the "wrong" sign. Another obvious but important point is that Poincar\'e duality is consistent with dimensional reduction and oxidation. Reducing over timelike directions, the forms may obtain additional minus-signs, but then the number $T$ also changes, and the Poincar\'e duality relation remains intact under dimensional reduction.
\end{enumerate}
In the dimensional reduction of a theory we find many form fields. Poincar\'e duality imposes relations between the signs of these form fields, and in this way also restricts us in our possibilities in assigning signs to various form fields. After having fixed a space-time signature and a definite sign for a form field, the sign for the dual form field is fixed. In the algebraic approach to these theories using $\tp{G}$ algebra's many signs cannot be chosen arbitrarily because the algebraic relations relate them. It turns out however that this is not the case for Poincar\'e duality: This is not automatically built in in $\tp{G}$-algebra's, but has to be imposed as a separate constraint. We will derive the precise form of the constraint in section \ref{Ggeneral}. 

\subsection{T-Duality}

From the effective field theory point of view, T-duality is a reinterpretation of the spectrum of scalar charges. Such charges couple to 1-form vector fields, so we focus on the sector of the effective field theory containing the 1-form vector fields. 

One of the sectors contributing to the 1-form sector is the gravity-sector. We reduce $\widetilde{D}$-dimensional general relativity to $D$ dimensions, and allow for the possibility that the $n= \widetilde{D} - D$ directions over which we are reducing may be timelike. The part of the Lagrangian describing the Kaluza-Klein-vectors is
\be
{\cal L}_{KK} = \sum_{i=1}^{n} \eta_i f_i(\phi_k)^{-2}\left( F_{\mu \nu, i} \right)^2 \label{KKLag}
\ee
De symbols $\eta_i = \pm 1$ indicate whether the direction corresponding to $i$ was a time-like ($\eta_i=-1$) or a space-like ($\eta_i=1$) one. The prefactor $f_i(\phi_k)$ depends on $i$ and on the dilatonic scalars $\phi_k$ in the theory. We also define the overall signature of the reduced dimensions $\eta$ ($T$ is the number of time directions among the directions over which one is reducing)
\be
\eta = \prod_{j} \eta_j=(-)^T.
\ee
In the most familiar reduction schemes \cite{Ortin:2004ms, Cremmer:1999du} (using "upper triangular" or "Borel" gauges) $f_i(\phi_k)$ is of the form $\exp\inp{\lambda_i}{\phi}$, but for reasons to be discussed later we will allow more general forms. The $F_{\mu \nu, i}$ obey Bianchi identities of the form
\be
\dif F_i = \textrm{derivatives of scalars} \wedge F_j
\ee
Again, the exact expressions are depending on the dimensional reduction scheme, but will not be of much importance to us.
 
Equally important, when given a sector in the theory with Lagrangian (\ref{KKLag}), then, under suitable conditions, we can reinterpret it as representing reduction over a number of directions with appropriate signature, a procedure that is known as oxidation.

Now assume that the Lagrangian contains a second part, coming from the reduction of the dynamics of a space-time $(n+1)$-form $C_{\mu_1, \ldots, \mu_{n+1}}$. We will generalize slightly, and actually allow for a multiplet of $p$ such $n+1$-forms. The reduction of such a $p$-plet of forms over $n$ dimensions gives rise to another sector in the reduced theory with 1-forms, with Lagrangian
\be
{\cal L}_{n} = \eta \sum_{i=1}^{p} \theta_i g_i(\phi_k)^{-2}\left( G_{\mu \nu; i_1 \ldots i_n;i} \right)^2 \label{KKn}
\ee
The signs $\theta_i = \pm 1$ are the signs appearing in front of the kinetic term for the $C_{(n+1),i}$-field before reduction.

Provided the $g_i$ and $f_i$ take similar forms (as they do in available dimensional reduction schemes), the Lagrangians (\ref{KKLag}) and (\ref{KKn}) are similar in form. They arose from the reduction of a $p$-plet of $n$-forms over $n$-dimensions, coupled to gravity, but it is tempting to explore the possibility that they can also be interpreted as coming from the reduction of an $n$-plet of $p$-forms, reduced over $p$ dimensions. The signature of the dual theory is set by the $p$ signs $\eta \theta_i$, while the $n$ signs $\eta_i$ will appear in front of the kinetic terms for $n$ $p$-forms. We also note that the existence of an $(n+1)$-form $C_{\mu_1, \ldots ,\mu_{n+1}}$ coupled to gravity allows for an extremal $n$-brane solution. It is to be expected that the above duality exchanges a $p$-plet of $n$-branes in one theory, for an $n$-plet of $p$-branes in the other theory.

An immediate condition that has to be met for this reinterpretation, is the presence of a subsector of the scalar sector that can be interpreted as an $SL(n)/SO(n-q,q) \times SL(p)/SO(p-r,r)$ sigma-model \cite{Julia:1980gr}. The $SL(n)/SO(n-q,q)$ factor arises automatically in the reduction of the gravity sector, where $q$ is the number of $\eta_i = -1$ in (\ref{KKLag}), encoding the number of time-like directions involved in the reduction. Similarly, $r$ is the number of $\theta_i = -1$ in (\ref{KKn}). The condition on the presence of the relevant sigma models however is only a strong one if $p > 1$, because when $p=1$ the $SL(p)$ factor trivializes. The possible presence of enhanced symmetries is however a significant point. Further dimensional reduction will not destroy these enhanced symmetries, and hence they can be equally well studied in theories that have been reduced further. In \cite{Keurentjes:2002xc, Keurentjes:2002rc} it was demonstrated how from enhanced symmetries in 3 dimensions the bosonic sector of the higher-dimensional theory can be reconstructed; it turns out no additional conditions are needed.

These considerations give a potential duality between two, \emph{a priori different} theories. In general the two dual theories do not even live in a space-time of the same dimension. We should stress however the "a priori": As we are focusing on a subsector of a theory to start with, the dual description also involves only a sector of a theory and extending to the full theory we may be studying different truncations of the same theory.

We illustrate these considerations with some familiar examples. The membrane of 11-dimensional M-theory dualizes to a doublet of strings in a dual 10 dimensional theory; of course this is IIB -supergravity. We reduce on a 2 torus, that can be of signature (0,2) or (1,1). With signature (0,2) $\eta_1 =\eta_2= \eta = \theta$, and the dual theory has signature (1,9) with two standard 2-form terms. With signature (1,1) we have $\eta = -1$. Together with the standard sign of the M-theory 3-form kinetic term, this means that we get a time-like direction back for the $T^{1,1}$ we have sacrificed, but the different signs of the $\eta_i$ imply that the 2-form field kinetic terms have opposite signs \cite{Hull:1998vg, Hull:1998ym}. Also the other relations between $M, M', M^*$ and $IIB, IIB', IIB^*$ theories can be analyzed along these lines.

A 3-brane in a space-time of signature $(1,9)$, as relevant to IIB-theory, but also to the theory with conjectured $\tp{E_7}$ symmetry, dualizes to a triplet of strings in 8 dimensions. With the various options for the signature, the space-time signature of the dual space is always (1,7), but the 2-forms coupling to the strings have kinetic terms with signs $(+++)$ or $(-++)$ depending on the choice of signature on the 3-torus. It is easy to see that they are mixed by $SL(3,\R)/SO(3)$ respectively $SL(3,\R)/SO(1,2)$. Notice that for IIB-theory, the invariance of the 3-brane under $SL(2,\R)/SO(2)$ implies that the dual strings should be inert under the $SL(2,\R)$ factor in the $SL(2,\R) \times SL(3,\R)$ U-duality group in 7 dimensions, as they are.

Another simple example is given by the doublet of 5-branes in IIB-theory. According to the above rules, they should give rise to a quintuplet of membranes in 7 dimensions, transforming as a quintuplet under $SL(5)$ and under $SO(4,1)$ or $SO(5)$ \cite{Hull:1998ym, Keurentjes:2004xx}. The M-theory 5-brane dualizes to the 5 strings in 7 dimensions that are the electro-magnetic duals to the membranes.
 
As a final cautionary remark we stress that one should verify that all necessary conditions are really met. As an absurd example of too naive analysis: 7-dimensional M-theory has a decuplet of 3-branes, but this does not imply the existence of a triplet of 10 branes in a 14-dimensional theory; there is no underlying $SO(10)$-symmetry. 

\subsection{T-symmetry}

Upon invoking an extra dimension in the reduction, one can return to the original theory without returning to the original parameterization: First one reduces over $n$ dimensions, and replaces these by the $p$ dimensions of the dual system; with the extra circle we now have $p+1$ circles, from which we choose another set of $p$ coordinates, and replace these by $n$ coordinates of the original theory. 

From the effective field theory point of view, this involves the following Lagrangian. It is obtained by focusing on the subsector of 1-forms of the theory of $\tilde{D}$-dimensional General Relativity, coupled to an $n$-form, and reduced over $n$-dimensions.
\bea
{\cal L} =  & &\sum_{i=1}^{p+1} \eta_i f_i(\phi)^{-2}\left( F_{\mu \nu, i} \right)^2 \\
& + \theta_G \eta & \sum_{i=1}^{p+1} \eta_i h_i(\phi)^{-2} H_{\mu \nu,i}
\eea
where the $F_{\mu \nu, i}$ are the $p+1$ field strengths corresponding to the Kaluza-Klein vector bosons, whereas the $H_{\mu \nu, i} = (p \ !)^{-1}\varepsilon_{i i_1 \ldots i_p}G_{\mu \nu,i_1\ldots i_p}$ correspond to the reduced form field modes. The sign $\theta_G$ encodes the sign in front of the kinetic terms of the $C$-field before reduction, and $\eta$ is again the product of the signs of the compact time-directions. As there are ${p+1 \choose p}=p+1$ ways of reducing the form, there are just as many $G_{\mu \nu, i}$ as $F_{\mu \nu , i}$. Provided suitable conditions on the $f_i$ and $h_i$ are met, there is a duality exchanging the $F$'s and the $G$'s.
The overall signature of the $F$'s and $G$'s are exactly identical, up to an overall sign set by $\theta_G \eta$. Hence, if $\theta_G \eta=-1$, the duality exchanging the $F$'s with the $G$'s is a signature changing one. We conclude that the dual theory has 
\be
\eta'_i= \eta \theta_G \eta_i; \qquad \eta'= \theta_G^{p+1} \eta^{p}; \qquad \theta_G'= \theta_G^p \eta^{p+1}.
\ee

It is a fact is that the existence of extremal $p$-brane solutions coupling to the $G$-fields requires them to have a world volume signature $-\theta_G$: If $\theta_G=1$ then the number of timelike directions incorporated will be $1 \ \mod \ 2$, whereas when $\theta_G =-1$ the number of timelike directions is $0 \ \mod \ 2$ \cite{Hull:1998fh}. Hence, if we choose $\eta$ to be in correspondence with the world volume signature of extremal $p$-branes the above duality will always result in some dimensions changing signature. But this then immediately implies that the corresponding $p$-brane can wrap the cycle over which we are reducing. In string theory it is often argued that the fact that a brane can wrap a cycle softens the potential singularity that arises if such a cycle shrinks to zero size. It would of course be most interesting to study time-like T-duality for more complicated reductions to see if this is a property that is general. 

The charges coupling to the $F$-fields are the momenta of
gravitational waves, while the charges of the $G$-fields are winding
modes (where it is understood that momenta and winding can be
space-like and time-like). The spacing of the momenta is small for
large volumes, while the spacing of winding modes is large (and vice
versa for small volumes). As the hypothetical duality interchanges
these, we have a standard "large volume-small volume" duality. For
theories which are conjectured to be described by non-linear
realizations of $\tp{G}$ algebra's, exact formula's for the volumes
can (in principle) be extracted with the techniques
used in \cite{Englert:2003zs}.

With a little manipulation we see that
\be
\theta_G' = \theta_G (\eta \eta').
\ee
The kinetic term will therefore change sign if the number of compact
time dimensions changes from even to odd or vice-versa under the
duality.

As illustrative examples we recall that 11-dimensional $M$-theory has
a membrane solution with world volume signature $(T,S) =(1,2)$. We then
partition the space-time signature $(1,10)$ in a part corresponding to
the membrane world volume and a transversal part, and T-dualize on a
3--torus with signature $(1,2)$:
\be
(1,10) = (1,2) + (0,8) \rightarrow (2,1) + (0,8) = (2,9)
\ee
Under this duality the number of compact time-directions changes from 1 to 2, so we should expect a sign change in front of the kinetic terms for the 3-form. M-theory also has a 5-brane solution, with world volume signature $(1,5)$. Now the above duality applied to the 5-brane implies
\be
(1,10) = (1,5) + (0,5) \rightarrow (5,1) + (0,5) = (5,6)
\ee
and the sign in front of the 6-form kinetic terms does not change (and, by Poincar\'e duality, neither does the sign for the 3-form term).

We have to pay special attention to the situation in 4 dimensions, where the Kaluza-Klein gauge fields have duals that are also vector fields. We can set up the same Lagrangian, but now $G$ is essentially $\df{F}$. If we ignore the complications caused by Chern-Simons terms, we could formally Poincar\'e dualize, to derive that $\epsilon_G=(-)^{T-1}$, where $T$ is now the number of time directions in the \emph{non-compact} directions. Hence a signature changing duality requires $T$ to be even, that is $0,2,4$. For General Relativity in signature $(p,q)$ there exist extremal solutions of the form $\R^{p-4,q} \times TN^{4,0}$, $\R^{p-2,q-2} \times TN^{2,2}$ $\R^{p-2,q-2} \times -TN^{2,2}$ $\R^{p,q-4} \times TN^{0,4}$, with $\R^{a,b}$ flat space of signature $(a,b)$, and $TN^{a,b}$ a Taub-NUT-solution of signature $(a,b)$ \cite{Barrett:1993yn, Hull:1998fh}. Again we are in the situation that for a signature changing duality we always have an extremal brane-solution that can wrap the compact directions. 

In this case a signature changing duality gives dual signature
\be
(1,10)= (1,6) + (0,4) \rightarrow (6,1) +(0,4) = (6,5)
\ee
It is tempting to speculate that the existence of M-theory variants
with signature $(9,2)$ and $(10,1)$ is evidence for objects with $10$
and $11$ dimensional world volumes, but of course the present methods
are not suitable for discussing these (dimensional reductions to 2 and
1 dimension would be needed, in which case we never would have vectors
at our disposal). A candidate 10-dimensional object (the
``$M9$''--brane) has appeared at various places in the literature.

\section{Sigma models and non-linear realizations} \label{tech}

In the previous section we found suggestive dualities on the basis of
an analysis of the spectrum of charges, that couple to
vector-fields. Compactifying and reducing over an additional
dimension, the symmetries of the vector fields become reflected in the
set of scalar fields in the theory. These typically form sigma-models
on symmetric spaces of the form $G/H$, where $G$ is a non-compact
group, and $H$ is a real form of the complexification of the maximal
compact subgroup in $G$. As dualities from the effective field theory
point of view are nothing but ambiguities in the oxidation procedure,
we should now analyze the oxidation of the sigma-model. Here we have
powerful methods at our disposal, rooted in the representation theory
of $G$. For finite dimensional $G$ the oxidation procedure is
mathematically rigorous \cite{Keurentjes:2002xc, Keurentjes:2002rc},
for infinite dimensional $G$ a level expansion gives the correct
result at lower levels, but is not understood at higher levels
\cite{Damour:2002cu, West:2002jj, Nicolai:2003fw,
  Kleinschmidt:2003mf}.

\subsection{Ambiguities from oxidation of the scalar sector}

The Lagrangian for a coset sigma model on $G/H$ can be written in the following form
\be \label{Lang}
\int \dif^d x \ \hlf\textrm{Tr}\left(FF-F\omega(F)\right).
\ee
Here $F$ is the $Lie(G)$-valued ($Lie(G)$ is the Lie-algebra of $G$), left-invariant one-form (of the form $V^{-1}(\dif V)$ with $V \in G$), and $\omega$ is the involution that leaves $Lie(H)$ invariant. For non-simple $G$ the Lagrangian decomposes in several terms, each of which has the same structure. For definiteness we will assume $G$ to be simple in the following.

If $G$ is finite-dimensional, then $\textrm{Tr}(FF)$ is the standard bilinear form on $Lie(G)$ and hence positive definite. If in addition $\omega$ is a Cartan-involution, then by definition the bilinear form $-\textrm{Tr}(F\ \omega(F))$ is positive definite \cite{Helgason}, and hence the group structure does not contribute minus signs to the Lagrangian. As $\omega(h)=h$ for $h \in Lie(H)$, and the definition of Cartan involution implies that  $\textrm{Tr}(hh)=\textrm{Tr}(h \omega(h)) < 0$ for $h \neq 0$, $H$ will be compact by Cartans first criterion. In all other cases, $H$ is non-compact, and the trace $\textrm{Tr}$ over the group indices is not positive definite.

The one-form $F$ transforms as a connection under $G$ transformations
\bd
F \rightarrow g^{-1}Fg + g^{-1}(\dif g),
\ed
but the Lagrangian is only invariant if $g \in H$. Hence the theory is invariant under \emph{global} $G$-transformations $V \rightarrow gV$ (as these leave $V^{-1}(\dif V)$ invariant), and local $H$ transformations $V \rightarrow V h(x)$, and describes a sigma-model on the coset $G/H$.

This form of the Lagrangian suffices for discussions in $d$-dimensions, but we actually need to relate this Lagrangian to higher dimensional theories. To make the link, one makes the field-content of the Lagrangian visible in a two-step procedure. 

The first step consists of fixing a way to represent $Lie(G)$.  This is necessary to do actual computations with the Lagrangian (\ref{Lang}), and more particular, to specify the action of $\omega$. As an example, one can describe an $SL(p+q,\R)/SO(p,q)$ by representing $SL(p+q,\R)$ by $(p+q)$-dimensional square matrices, and $\omega$ by $\omega(F) = -\eta F^T \eta$, with $\eta$ an $SO(p,q)$ invariant metric and $F^T$ the matrix-transpose of $F$. However, such a representation is only convenient if not only $SO(p,q)$, but also $SL(p+q,\R)$ can be kept manifest in the oxidation. To also be able to describe the cosets for arbitrary $G$, we need a general prescription. It is then more convenient to work with the adjoint representation. We will fix a triangular decomposition $Lie(G) = {\cal H} \oplus n_- \oplus n_+$, where $\cal H$ is a fully reducible abelian subalgebra (a Cartan sub-algebra) and $n_\pm$ are composed of ladder operators associated to positive, and negative roots respectively \cite{Kac:1990gs}, and work with this. 

The second step consists of (partially) fixing the $H$-invariance of the theory. Essentially one is picking a set of variables to describe the coset $G/H$, after which (some) redundant coordinates are eliminated, but also (part of the) $H$-invariance is no longer manifest. This second step is needed to relate to the outcome of some dimensional reduction procedure, or in the case of the conjectures relating to $G^{++}$ and $\tp{G}$-algebra's \cite{Damour:2002cu, West:2002jj, Nicolai:2003fw, Kleinschmidt:2003mf}, to some reconstruction algorithm. One can leave some, or even all of the $H$-invariance intact, provided one relates to a formulation of the higher dimensional theory with manifest $H$-invariance, and uses a dimensional reduction algorithm that respects these.  One can for example reduce a theory from $D$ to $D'\geq 3$ dimensions, while keeping the local $SO(D-D')$ covariance of the vielbein-formalism manifest. We also note the existence of formulations of maximally supersymmetric theories with local $H$-invariance \cite{Hinv}.

We are emphasizing these two distinct steps, because when $H$ is non-compact, and hence when time-like T-dualities are possible, there are subtleties concerning the second step, the fixing of the $H$-invariance. But as $H$ is a local (gauge) symmetry, nothing can depend on the details of how we treat the $H$-invariance. We will nevertheless clarify the nature of the subtleties in our next subsection.

For the discussion of possible ambiguities in the interpretation of the subsector of the Lagrangian (\ref{Lang}) as coming from a higher dimensional theory, we only need to discuss $Lie(G)$. A first ambiguity arises because we have to fix a "gravitational subalgebra" of the form $A_n$, and there may be multiple non-equivalent embeddings (not related by conjugation with group elements) of these. This results in general in higher dimensional theories with different matter content, and lies at the heart of the algebraic description of T-duality.

Within a class of equivalent embeddings there are still ambiguities. The triangular decomposition is not invariant under the action of $G$, but there is a discrete subgroup of $G$, effectively the lift to $G$ of the Weyl group of $Lie(G)$, that normalizes the Cartan subalgebra, and also $n_+ \oplus n_-$, the set of ladder operators. It therefore permutes the fields associated to particular (combinations of) generators, and the oxidation of the theory to higher dimensions is ambiguous because of these permutations. This is what has been called T-symmetry. 

We note that, apart from the Weyl group, outer automorphisms of the algebra permuting the vectors of the root lattice will also result in ambiguities in the oxidation. These result however in gravitational sub-algebra's that are not conjugate (as the automorphism is outer) and hence this belongs to T-duality.

Even if part of the $G$-symmetry is kept, similar arguments apply. For
example, in the example sketched in the above where $SL(p+q,\R)$ (and
$SO(p,q)$) is kept, this group acts (in the fundamental, respectively
vector- representation) on some carrier space for the
representation. The assumption that this space is a $(p+q)$-torus,
makes backgrounds related by $SL(p+q,\R)$-transformations
inequivalent, but there is a subgroup of $SL(p+q,\R)$  with a rather
trivial action: it only permutes the basis-elements of the
carrier space. This permutation action is generated by the Weyl group
of $SL(p,q)$, and hence we are back in the previous situation.

\subsection{Subtleties}

For cosets $G/H$ where $H$ is non-compact, there are a number of subtleties compared to the case where $H$ is compact. We will highlight two issues that may seem confusing, but should not distract the reader from the main point. We think however that especially the first point deserves more attention than it has received thus far.

\subsubsection{The Borel gauge is a "bad" gauge for non-compact $H$}

In the dimensionally reduced theories sigma-models on symmetric spaces make their appearance. These symmetric spaces are of the form $G/H$, with $G$ a non-compact Lie-group. For the cases where the dimensional reduction/compactification includes only space-like circles, the group $H=H_c$ is the maximal compact subgroup of $G$; in the case time-like directions are included the compact group $H_c$ has to be replaced by a non-compact $H_{nc}$ with the same complexification as $H_c$.

For the sake of concreteness, we will work with $G$ that are split real forms, though all remarks generalize to $G$'s in other non-compact real forms. The Lie algebra ${\cal L}_G$ associated to $G$ allows a triangular decomposition in terms of a Cartan-subalgebra with generators $h_i$, and raising/lowering operators $e_{\alpha}$. The compact subgroup $H$ is now generated by generators of the form $e_{\alpha} - e_{-\alpha}$. 

This description of the group $H$ makes a very explicit use of the triangular decomposition. It allows the well-known action of the Weyl group on $G$, that will automatically act on $H$. 

To describe the coset $G/H$, one chooses a single representative of each orbit of $H$ on $G$. In physical language, this amounts to fixing the gauge. If we then act with the Weyl group on $G/H$ the gauge that we have chosen is typically not respected. Because the sigma-model is invariant under the left action of $H$ one can allow for an $H$-transformation that restores the gauge. The combination of a Weyl-reflection together with its gauge-restoring $H$-transformation then gives the symmetry transformation of various fields.

The most popular gauge-choice is the so-called Borel gauge, which relies on the Iwasawa-decomposition. The Iwasawa decomposition theorem states that any element $g$ of a non-compact group $G$ can be written as
\be
g = k \cdot a \cdot n
\ee
where $k$ is an element of the maximal compact subgroup in $G$, $a$ is an element of the Abelian group obtained by exponentiating (a subset of) the generators of the Cartan-subalgebra (generating non-compact symmetries), and $n$ an element of a unipotent group. In the special case that $G$ is a so-called split real form, the elements $a$ come from exponentiating all of the Cartan-subalgebra, and the elements $n$ are obtained by exponentiating elements from the step-generators associated to positive roots.

It is clear that this decomposition is of a great value for the description of cosets $G/H$ where $H$ is the maximal compact subgroup of $G$. In that case the elements of $G/H$ can be written as
\be \label{Borel}
g = \exp \left(\sum \phi_i h_i \right) \exp\left(\sum_{\alpha \in \Delta^+} C_{\alpha} e_{\alpha} \right),
\ee
$\Delta^+$ being the set of positive roots. In this gauge the action of $H$, generated by the algebra-elements $e_{\alpha} - e_{-\alpha}$ is completely fixed. A caveat in the application of the Iwasawa decomposition theorem is, in the context of the applications discussed here, that it is proven for finite-dimensional $G$ only. The crucial thing to establish is that \emph{any} element of $G$ can be written as in the decomposition, and it seems that this needs the relation between the Lie-algebra, and Lie-group, which is not well understood for general (not necessarily finite-dimensional) Kac-Moody algebra's. It is however generally assumed that an infinite-dimensional analogue of the Iwasawa decomposition exists \cite{Damour:2002cu}. 

More relevant in the situations of interest to us is that $H$ is not always compact. It was observed in \cite{Hull:1998br} that the relevant $H$ is always a real form of the complexification of the maximal compact subgroup $H$, but that it is not necessarily the compact real form. These forms can be described by replacing some of the antihermitian generators $e_{\alpha} - e_{-\alpha}$ by hermitian generators $e_{\alpha} + e_{-\alpha}$ \cite{West:2003fc, Englert:2003py, Schnakenburg:2003qw, Keurentjes:2003hc}. Demanding closure of the algebra leads to tools that allow to classify the relevant possible denominator subgroups \cite{Keurentjes:2004xx}.

By replacing a compact $H$ by a non-compact one, we are however
outside the scope of the Iwasawa decomposition theorem. If one
nevertheless insist on the usage of elements of the form (\ref{Borel})
we have to pay a price: Of course these elements are contained in the
coset, but the reverse is no longer true, not every element in the
coset can be written in the form (\ref{Borel}). This can be
immediately seen in examples, but also directly. The coset $G/H$,
where $H$ is not a maximal compact subgroup, must contain compact
cycles (because there must be compact orbits in $G$ that cannot be
compensated with $H$-transformations). But the ranges of the variables
in (\ref{Borel}) are non-compact, and consequently, the variables
(\ref{Borel}) describe only an open proper subset of $G/H$.

The prototype example is found in the space $SL(2,\R)/SO(1,1)$. The Borel gauge essentially amounts to the "upper triangular gauge"; elements in this gauge have the lower left entry set to zero. Consider now the element
\be
\left(\ba{cc}
\cos \phi & -\sin \phi \\
\sin \phi & \cos \phi 
\ea \right)
\in SL(2,\R)
\ee
If we try to bring this to upper triangular form by multiplying on the left by an element of $SO(1,1)$, it requires essentially solving the equation
\be
\tan \phi = \pm \tanh \omega 
\ee
for some $\omega$. But it is immediately clear that this cannot have solutions when $|\tan \phi| \geq 1$. This is in a nutshell the problem for any compact cycle in $G/H$.

Of course one can define the set $S$ of elements $k \cdot a \cdot n$ for all $k,a,n$, with $k \in H$. This set $S$ is in general not a group, because it will not close under group multiplication. Instead $S$ is a proper subset of $G$. Hence the set of all representatives $a \cdot n$ will only cover part of $G/H$. Whether this is a serious problem depends on the application we have in mind. We argued that T-dualities are generated by a discrete set of transformations on $G$. Now if we start with $S$ instead of $G$, then the image of $S$ under such transformations is generically not contained in $S$. In explicit language, the transformation that brings us back to our gauge choice does not exist! Hence in the end we can only do an analysis on
\be
\bigcap_{w \in W} \  \textrm{Im}_w(S)
\ee
where $W$ contains all the transformations (Weyl reflections, outer
automorphisms) we are interested in. For an extension to all of $G/H$
one has to rely on some sort of continuation procedure. It would be
much simpler to not rely on the Borel gauge for this kind of
computation at all, but it seems however that a general "good gauge
choice" (a gauge that can always be imposed) is unavailable even for finite-dimensional $G/H$, not even
mentioning infinite-dimensional $G$'s \footnote{Good gauge choices are known in specific situations. For example one can prove that for $SL(2,\R)/SO(1,1)$, with the $SL(2,\R)$ elements written as 
\bd 
\left( \ba{cc} a & b \\ c & d \ea \right), \qquad a,b,c,d \in \R , \ ad-bc =1,
\ed
the gauge condition $ac+bd =0$ can always be established. But we are
unaware of a general, good
(and useful !) gauge fixing procedure for general $G/H$.}. Luckily, as far as T-dualities are concerned, we do
not need to rely on any gauge choice. 

Moreover, we will argue later that in the context of $\tp{G}$
algebra's the use of compact subgroups in the coset construction is
inconsistent with Poincar\'e duality. But then the Borel gauge, used
throughout the literature thus far, misses part of the compact cycles
in $\tp{G}/K(\tp{G})$, and there are as a matter of fact infinitely
many independent such cycles if $K(\tp{G})$ is not compact!

\subsubsection{The action is not invariant under the Weyl-group}

A less important problem, which may nevertheless be confusing is the fact that the action (\ref{Lang}) is generically not invariant under the transformations we are considering. This can be seen easily already if we restrict to those duality transformations that are generated by Weyl reflections. These can be implemented by conjugation with an element $W$, acting as
\be
W \exp{g} W^{-1} = \exp{ w(g) } \qquad g \in \textrm{Lie}(G)
\ee
The element $W$ is an element of the maximal \emph{compact} subgroup in $G$. In the action (\ref{Lang}) the one form $F$ transforms as $F \rightarrow WFW^{-1}$. The action is then invariant if
\be \label{omegainv}
\omega_W(F) \equiv \omega(WFW^{-1}) = W \omega(F) W^{-1} \quad \forall W, F
\ee
We may analyze this equation as follows: To every element $g$ of the algebra there is a one parameter-subgroup of $G$ given by $\exp tg$. The Lie algebra homomorphism $\omega$ lifts to a group homomorphism $\Omega$ by defining $\Omega(\exp tg) = \exp(t \omega(g))$. Though not every element of $G$ can be written as $\exp tg$, any element can be written as a product of such elements, and then the homomorphism property guarantees that $\Omega$ can be extended to the whole group. We then compute
\be
e^{\omega(WFW^{-1})} = \Omega(e^{WFW^{-1}}) = \Omega(W e^F W^{-1}) = \Omega(W) e^{\omega(F)} \Omega(W^{-1}) 
\ee
The condition (\ref{omegainv}) can then be rewritten to
\be
\left(W^{-1} \Omega(W)\right) e^{\omega(F)} = e^{\omega(F)} \left(W^{-1} \Omega(W) \right)
\ee 
Because this has to valid for any $F$, this implies that $W^{-1} \Omega(W)$ is proportional to the identity (by Schurs Lemma). This is trivially the case if $\Omega(W) =W$, that is if $W \in H$. However $W$ is in the maximal compact subgroup of $G$, and this does not coincide with $H$ if $H$ is non-compact. Because $W$ is in the compact subgroup of $G$, we could also have $W^{-1}\Omega(W)=\lambda$, with $\lambda$ a phase , but this is generically not the case. So for non-compact $H$, the action is generically not invariant under Weyl-reflections.

This should not confuse the reader. It will be clear that also
$\omega_W(F)$ defines a formulation of the sigma-model on
$G/H$. Different $\omega_W$ do not in any way affect $G$, but only
change the way we embed $H$ in $G$. In view of the ambiguities we are
interested in, we should not view models defined by different
$\omega_W(F)$ as different models, but as different realizations of
the same model. If one could encode $\omega$ in a way that would not
rely on a specific basis for the algebra or a specific realization,
one might avoid the changes in the Lagrangian under Weyl reflections;
such invariant Lagrangians appear not to have been constructed, and it
is not even clear (to the author) that they exist.

\section{$\tp{G}$ algebra's: General theory} \label{Ggeneral}

We now explicitly turn to $\tp{G}$-algebra's. The present section is devoted to $\tp{G}$-algebra's in general, in the next section we will do computations with explicit $\tp{G}$ algebra's.

\subsection{Defining $\tp{G}$ and $K(\tp{G})$}

The $\tp{G}$-algebra's are essentially defined by their Dynkin diagrams. The Dynkin diagram for a simple group $G$ consists of $n$ nodes, where $n$ is the rank, which we label by an integer $i$ ranging from $1$ to $n$. To this diagram we add three more nodes, labelled by $0$, $-1$ and $-2$. The node labeled by $0$ is connected to the Dynkin diagram to obtain the extended Dynkin diagram, also known as the Dynkin diagram of the untwisted affine Lie algebra $G^+$ associated to $G$. We then connect the node labeled by $-1$ with a single line to the node $0$, to obtain the diagram of the overextension or canonical hyperbolical extension $G^{++}$ of $G$. Finally, the node $-2$ is connected by a single line to the node $-1$ to obtain the diagram of the triple extended algebra $\tp{G}$. Via this node by node extension, there is a natural embedding 
\be
\tp{G} \supset G^{++} \supset G^+ \supset G
\ee
We will make use of some results for $G^+$ and $G$ later.

From this diagram the Cartan matrix $A=(a_{ij})$, with $i,j$ in the index set $I \equiv \{0, 1, \ldots
10 \}$, may be reconstructed by the standard procedure \cite{Kac:1990gs}. The construction of the algebra $\tp{G}$ starts by choosing a real vector space ${\cal H}$ of dimension $n+3$ and linearly independent sets $\Pi = \{ \alpha_{-2}, \ldots,\alpha_{n} \} \subset {\cal H}^*$ (with $\cal H^{*}$ the space dual to ${\cal H}$) and $\Pi^{\vee} = \{ \alpha_{-2}^{\vee}, \ldots,\alpha_{n}^{\vee} \} \subset {\cal H}$, obeying $\alpha_j(\alpha_i^{\vee}) = a_{ij}$. The $\alpha_i$ are called the simple roots of $\tp{G}$, and $\alpha_i^{\vee}$ are the associated simple coroots.

From the Cartan matrix the algebra $\tp{G}$ can be constructed. The generators of the algebra consist of $n+3$ basis elements $h_i$ of the Cartan sub algebra ${\cal H}$ together with 22 generators $e_{\alpha_i}$ and $e_{-\alpha_i}$ ($i \in I$), and of algebra elements obtained by taking multiple commutators of these. These mutual commutators are restricted by the algebraic relations:
\be
[h_i,h_j]=0 \qquad [h_i, e_{\alpha_j}] = \alpha_{j}(h_i) e_{\alpha_j} \qquad [h_i,f_{-\alpha_j}] = -\alpha_j(h_i) e_{-\alpha_j} \qquad [e_{\alpha_i}, e_{-\alpha_j}] =\delta_{ij} \alpha_i^{\vee}h_i;
\ee
and the Serre relations
\be
\textrm{ad}(e_{\alpha_i})^{1-a_{ij}} e_{\alpha_j} =0 \qquad \textrm{ad}(e_{-\alpha_i})^{1-a_{ij}}e_{-\alpha_j} =0 
\ee

There is the root space decomposition with respect to the Cartan subalgebra, $\tp{G} = \oplus_{\alpha \in {\cal H}^*} g_{\alpha}$, with
\be
g_{\alpha} = \{ x \in \tp{G}: [h,x] = \alpha(h) x \ \forall \ h \in {\cal H} \}
\ee
The set of roots of the algebra, $\Delta$, are defined by
\be
\Delta = \{ \alpha \in {\cal H}: g_{\alpha} \neq 0, \alpha \neq 0  \}
\ee
The root lattice $P(\tp{G})$ consists of all linear combinations of roots with integer coefficients. The roots of the algebra form a subset of the root lattice $\Delta \subset P$. We will denote the generators of $g_{\alpha}$ by $e^k_{\alpha}$, where $k$ is a degeneracy index, taking into account that the dimension of $g_{\alpha}$ may be bigger than 1. If $\dim(g_{\alpha}) =1$, we will drop the degeneracy index, and write $e_{\alpha}$ for the generator. This in accordance with previous notation, as $\dim(g_{\alpha_i}) =1$ for $\alpha_i$ a simple root. By using the Jacobi identity, one can easily prove that $[e^i_{\alpha}, e^j_{\beta}  ] \in g_{\alpha+\beta}$, if this commutator is different from zero.

A particular real form of the $\tp{G}$ algebra is the split real form. The Lie-algebra of the split real form consists of linear combinations of the generators constructed thus far with real coefficients. Almost all of the present literature on $\tp{G}$ algebra's is devoted to split real forms. Other real forms can be constructed by defining realizations for the Cartan involution. It is easy to see that for 3 dimensional coset sigma models involving non-compact real forms of $G$ (see \cite{Keurentjes:2002rc} for an exhaustive list) there exist corresponding $\tp{G}$ algebra's defined by extending the Satake-Tits diagram in the obvious way. We will however also restrict mostly to split real forms.

Using the standard bilinear form on the root space \cite{Kac:1990gs} we define the simple coweights (which are elements of ${\cal H}$) by
\be
\inp{\alpha_i}{\omega_j} = \delta_{ij}.
\ee
The coweight lattice, which we call $Q(\tp{G})$ consists of linear combinations of the fundamental coweights with coefficients in $\Z$, and is the lattice dual to the root lattice $P(\tp{G})$. 

The Weyl group $W(\tp{G})$ of $\tp{G}$ is the group generated by the Weyl reflections in the simple roots
\be
w_i(\beta) = \beta - 2 \frac{\inp{\alpha_i}{\beta}}{\inp{\alpha_i}{\alpha_i}} \alpha_i
\ee
The Weyl group leaves the inner product invariant
\be
\inp{w(\alpha)}{w(\beta)}=\inp{\alpha}{\beta} \qquad w \in W(\tp{G})
\ee 

In \cite{Keurentjes:2004bv, Keurentjes:2004xx} we demonstrated that the closure of the group $K(\tp{G})$ demands that the hermiticity properties of its generators can be conveniently encoded in an element $f \in Q(\tp{G})$. The group $K(\tp{G}$) is then generated by the generators
\be
e^k_{\beta} - e^{i \pi \inp{\beta}{f}} e^k_{-\beta} \quad \in \tp{G}
\ee
From the definitions it is straightforward to derive that $f$ and $f+ 2q$ define the same group for $q \in Q(\tp{G})$. As a consequence $f$ can be expanded in the fundamental coweights as 
\be
f = \sum_{i= -2}^n p_i \omega_i, \qquad p_i \in \Z_2 = \{0,1 \}
\ee
These definitions reduce the characterization of the real form of the
infinite-dimensional group $K(\tp{G})$ to the specification of the
$n+3$ $\Z_2$-valued coefficients $p_i$.

The coset $\tp{G}/K(\tp{G})$ encodes the properties of a class of physical theories, along with all their reductions and oxidations. To recover the field content of a $D$-dimensional theory one chooses a regular gravitational subalgebra $SL(D,\R)$ and decomposes under this algebra. The representation content at lower levels consist of the gravitational sector, a scalar sector transforming under the centralizer of $SL(D,\R)$ in $\tp{G}$, and a bunch of form fields, the higher levels are not understood \cite{Damour:2002cu, West:2002jj, Nicolai:2003fw, Kleinschmidt:2003mf}.

The space-time signature in turn is encoded in how the $SL(D,\R)$ intersects with $\tp{G}$, as its intersection is the algebra $so(p,D-p)$ \cite{Keurentjes:2004bv}. This $so(p,D-p)$ algebra at level 0 implies that the form fields at higher levels transform as $so(p,D-p)$-tensors. There is however the possibility of additional signs (because $\tp{G}$ is strictly bigger that $so(p,D-p)$) which was interpreted in \cite{Keurentjes:2004bv} as a consequence of the freedom to multiply the kinetic terms of the tensor fields with additional signs.

We have argued thus far that Poincar\'e duality restricts us in our ability to add signs to the algebra in a consistent way. In the next section we will derive how the restrictions from Poincar\'e duality should be consistently built in into the above framework.
 
\subsection{Poincar\'e duality and the $i_0$ criterion}

At first it may seem that for general $\tp{G}$ we can only do computations on a case-by-case basis, as the matter content depends sensitively on the algebra.  All of theories described by $\tp{G}$ algebra's however have a universal feature: if reduced to 3 dimensions, they describe a scalar sigma model with global $G$ symmetry. The duals to the axionic scalars are vectors in 3 dimensions, which should also transform under $G$. These matter fields can be recovered from the decomposition relevant for 3 dimensions: 
\be
\tp{G} \rightarrow SL(3) \times G.
\ee
This decomposition automatically implies a level expansion, where "level" is defined with respect to the $\alpha_0$-node.
It is relatively easy to verify that the first two levels with respect to this decomposition contain the representations
\be
\begin{tabular}{|c|c|}
\hline
level & $sl(3) \oplus G$ irrep \\
\hline
0 & $\mathbf{(8,1) \oplus (1,\textrm{\bf dim}(G))}$ \\
1 & $\mathbf{(3, \textrm{\bf dim}(G))}$ \\
\hline
\end{tabular}
\ee
The interpretation of the matter spectrum is straightforward. At level 0 the $\mathbf{(8,1)}$ is the adjoint of $SL(3)$, which in turn is part of the $GL(3)$ of 3-dimensional General Relativity; these are singlets with respect to $G$. The irrep $\mathbf{(1,\textrm{\bf dim}(G))}$  however forms a singlet under the $SL(3)$ of General Relativity, hence these are scalars, and there are $\textrm{dim}(G)$ of them, enough to fill the adjoint of $G$\footnote{When fixing a gauge, almost half of these are eliminated. Remember however that we have chosen not to fix the gauge.}. The first level consists of vectors (the $3$-dimensional irrep of $SL(3)$), formally dual to the axions of the scalar sector, and hence also transforming in the adjoint of $G$.

Hence, on the levels 0 and 1, we have 4 copies of the root system of
$G$, a singlet of $SL(3,\R)$ at level 0 for the scalar sector, and a triplet of $SL(3)$ at level 1, for the vectors. If we denote by $\alpha_H$ the highest root of $G$, embedded in the obvious way in the $\tp{G}$ root system (by identifying $\alpha_i$ with $i > 0$ as the simple roots of $G$), the corresponding roots can be written as 
\be
\begin{tabular}{|c|cc|}
\hline
level & \mc{2}{|c|}{$\tp{G}$-roots} \\
\hline
0 &  0 & $\pm$ positive roots of $G$ \\
\hline
  & $\alpha_0 + \alpha_H$ & $\pm$ positive roots of $G$ \\
1 & $\alpha_{-1} + \alpha_0 + \alpha_H$ & $\pm$ positive roots of $G$\\
  & $\alpha_{-2} + \alpha_{-1} + \alpha_0 + \alpha_H$ & $\pm$ positive roots of $G$ \\ 
\hline
\end{tabular}
\ee

Now we turn back to the physical 3-dimensional theory. If the space-time signature of the 3-dimensional space-time is $(+++)$ (3 spatial dimensions) then the overall signs of the 3 components of the vector should be $(---)$; Poincar\'e duality requires an overall minus sign in front of the vector. If the space-time signature is $(---)$ (3 time-dimensions), then the signs again have to be $(---)$, because now Poincar\'e duality tells us that there is no extra sign for an odd number of time dimensions. Similarly, for signature $(-++)$ the vector has $(-++)$ while for signature $(--+)$ the vector has $(++-)$. Summarizing, the number of relative minus signs appearing in the vector is always odd, regardless of the signature. 

We can read off the relative signs for the vector from the algebra, by computing:
\be \label{vecsigns}
\inp{\alpha_0+ \alpha_H}{f} \mod \ 2, \quad \inp{\alpha_{-1} + \alpha_0 + \alpha_H}{f} \mod \ 2, \quad \inp{\alpha_{-2} + \alpha_{-1} + \alpha_0 + \alpha_H}{f} \mod \ 2
\ee
We have just derived that Poincar\'e duality requires that there is an
odd number of minus signs among these 3 signs. This can be easily
checked by summing the 3 relative signs modulo 2. We should therefore
demand the following

{\bf Consistency criterion:} The function $f$ specifies a $K(\tp{G})$ consistent with Poincar\'e duality if and only if
\be \label{crit}
i_0(f) \equiv \inp{\alpha_{-2} + \alpha_0 + \alpha_H}{f} \ \mod \ 2 =1.
\ee
 
With this criterion, the signs of the vector become
\be
\inp{\alpha_{-2}}{f} + 1 \ \mod \ 2; \qquad \inp{\alpha_{-2} + \alpha_{-1}}{f} + 1 \ \mod \ 2; \qquad \inp{\alpha_{-1}}{f} + 1 \ \mod \ 2.
\ee
As the Lorentz group is encoded in the generators as 
\be
e_{\beta} - \exp{ i \pi \inp{\beta}{f}} e_{-\beta}, \qquad \beta = \alpha_{-1}, \alpha_{-2}, \alpha_{-1} + \alpha_{-2},
\ee
it is straightforward to check that the requirement (\ref{crit}) is sufficient to obtain the right signs. 

The derivation of the criterion was done for the 3-dimensional theory. We noted however that Poincar\'e duality is consistent with reduction and oxidation, which are automatically implemented in the algebra. The criterion (\ref{crit}) should therefore be the correct criterion in \emph{any} dimension for space-time, even though it is much less straightforward to derive it for the more general case.

The above expression for $i_0(f)$ is a universal one, but when we wish to apply it to a specific $\tp{G}$ algebra, we need the, algebra dependent, expression for $\alpha_H$. For any $G$ $\alpha_H$ is the so-called "highest root", and allows an expansion in the simple roots as
\be
\alpha_H = \sum_{i=1}^r a_i \alpha_i
\ee
The integer numbers $a_i$ are known as the root integers, and because of the $\mod \ 2$ nature of the $i_0$ criterion, we only need their values modulo 2. The reader can find them for example in \cite{Kac:1990gs}, table Aff 1 in chapter 4. Later we will discuss the separate $\tp{G}$ algebra's explicitly, and give the explicit expressions for $i_0(f)$ for all the cases. 

\subsection{Properties of $i_0$}

The quantity $i_0(f)$ enjoys the following two important properties
\bea
i_0(f) & = & i_0(w(f)) \qquad \forall w \in W(\tp{G}); \label{Weylinv}\\
i_0(f) & = & i_0(f + 2 q) \qquad \forall q \in Q(\tp{G}). \label{coweightinv}
\eea

We will call $i_0(f)$ a "mod 2 index" of $K(\tp{G})$. Although $i_0(f)$ is formulated in terms of $f$, the above statements indicate that $i_0$ not so much characterizes $f$, but rather the equivalence classes of $f$ under the Weyl group and shifts over twice the coweight lattice. These transformations do not alter $K(\tp{G})$, but only its embedding in $\tp{G}$, and in that sense $i_0$ is a quantity that contains information on $K(\tp{G})$. We will now prove the invariance statements (\ref{Weylinv}) and
(\ref{coweightinv}).

The statement (\ref{Weylinv}) states that $i_0(f)$ takes the same values on $f$'s related by Weyl-reflections. It is convenient to use the orthogonality properties of the Weyl reflections with respect to the bilinear form
\be
\inp{\beta}{w(f)} = \inp{w^{-1}(\beta)}{f}
\ee
We then prove (\ref{Weylinv}) as follows: $\alpha_0 + \alpha_H$ is an imaginary root of $G^+$. Imaginary roots of affine Kac-Moody algebra's are all null $|\alpha_0 + \alpha_H| = 0$ \cite{Kac:1990gs}, and orthogonal to any real root of the algebra, which in this case are the roots $\alpha_0, \ldots, \alpha_r$. Together with the fact that inspection of the Dynkin diagram teaches us that $\alpha_{-2}$ is orthogonal to $\alpha_0, \ldots, \alpha_r$ (by the defining properties of the Dynkin diagram), this shows that (\ref{crit}) is invariant under Weyl reflections in $\alpha_0, \ldots, \alpha_r$. It remains to check for the Weyl reflections in $\alpha_{-1}$ and $\alpha_{-2}$. On the basis of elementary roots these have the following effect:
\be
\ba{l@{\ : \ }lll}
w_{\alpha_{-1}} &  \alpha_{-2} \rightarrow {\alpha_2} + \alpha_{-1}; & \alpha_{-1} \rightarrow \alpha_{-1}; & \alpha_0 \rightarrow \alpha_0 + \alpha_{-1}; \\
w_{\alpha_{-2}} &  \alpha_{-2} \rightarrow {-\alpha_2}; & \alpha_{-1} \rightarrow \alpha_{-1} + \alpha_{-2}; & \ea  
\ee
and roots that are not listed remain invariant. It is immediately verified that $w_{\alpha_{-1}}$ and $w_{\alpha_{-1}}$ also leave $i_0(f)$ invariant, hence all generators of the Weyl group leave $i_0(f)$ invariant. This last statement implies that any element of the Weyl group leaves $i_0(f)$ invariant.

If the diagram of $\tp{G}$ admits a diagram automorphism, the algebra of $\tp{G}$ admits an outer automorphism (that actually is an extension of the outer automorphism of $G$). In the cases where these appear ($\tp{A_n}$, $\tp{D_n}$, $\tp{E_6}$), it can be verified (see the explicit expressions given later) that the index $i_0(f)$ is not only invariant under the Weyl group, but also under the outer automorphism. In essence this can be directly traced to the fact that the set of root integers $a_i$ is invariant under outer automorphisms \cite{Kac:1990gs}.

The second statement, that $i_0(f)$ takes the same value on $f$ and $f+ 2 Q(\tp{G})$ follows trivially from the $\mod \ 2$ property of $i_0(f)$ and the definition of $Q$.

These properties are crucial for the consistency of the framework. We have claimed that Weyl reflections and outer automorphisms in the $\tp{G}$ algebra correspond to a kind of duality group, acting on the physical theories constructed from the algebra, and that theories not satisfying the $i_0$-criterion are incompatible with Poincar\'e duality. It is important then that Weyl reflections should respect Poincar\'e duality. The properties we have just verified ensure that they indeed do.

An immediate consequence of the criterion (\ref{crit}) is that it is
inconsistent with the use of $f=0$, that is with a "compact"
subalgebra. This means that within this framework, even if we decided
to work with Euclidean theories, we will always encounter minus signs
and hence non-compact denominator subgroups! Also Wick rotations must
be performed with care, it seems that at least for some applications
we should demand consistency with the $i_0$ criterion. Then, although
the denominator subgroup $K(\tp{G})$ can be altered by such a
procedure, we can never change it to the compact algebra. Of course
this could have been anticipated: For a Euclidean theory reduced to 3
dimensions, Poincar\'e duality implies that there should be a relative
minus sign between 3-dimensional scalars and the dual vectors, as has
been known for long \cite{Gibbons:1979xm}.

For theories in which (anti-)self-dual tensors occur, we have to require specific signatures for space-time. If we do not, there is a set of tensors in the level decomposition of the $\tp{G}$- algebra that does not have duals (namely the would-be self-dual tensors that cannot be self-dual because it is inconsistent with the space-time signature). Imposing the $i_0$ criterion, Poincar\'e duality is implemented correctly, and this problem cannot occur. We will check later in explicit examples that everything works out as it should. 

\subsection{Other "mod 2 indices"}

The existence of the Weyl-invariant index $i_0(f)$ immediately evokes the question: Do there exist other $\mod \ 2$, Weyl-invariant quantities? It is clear that if they do, these could be used to distinguish between the various orbits of the Weyl group, and extract information on the theory without fixing a level decomposition.

It seems there are no further general indices of the type of $i_0(f)$, but a survey among specific $\tp{G}$ algebra's  will show that some of them admit other indices.  It also turns out that though they do distinguish between the orbits of the Weyl groups, they are not "fine enough" in the sense that it is not possible to tell every orbit apart from others by comparing these "indices".

The requirement of invariance under shifts by $2Q_{\tp{G}}$ is automatically built in if we use the following ansatz for a "$\mod \ 2$ index":
\be
i(f) = \inp{\sum_{i=-2}^r p_i \alpha_i}{f} \mod \ 2 \qquad p_i \in \Z
\ee
The $\mod \ 2$ property implies that we can restrict $p_i$ to the values $0,1$ of the additive group $\Z_2$. 

The question is now which combinations of $p_i$ make $i(f)$ invariant under Weyl-reflections. A trivial solution consists of setting all $p_i=0$, and the existence of $i_0(f)$ demonstrates that there is a non-trivial solution. An easy way to construct all non-trivial indices goes as follows. One can encode a candidate $i(f)$ by writing the Dynkin diagram of $\tp{G}$, and label all nodes $i$ with the number $p_i$. The action of a Weyl reflection in the simple root labeled by node $i$ can be computed to be
\be
w_i( \sum_j p_j \alpha_j ) = \sum_j (p_j - 2\frac{\inp{\alpha_j}{\alpha_k}}{\inp{\alpha_j}{\alpha_j}} p_k) \alpha_j 
\ee
As we will be computing modulo 2, the only property of the Cartan matrix we are interested in is whether its entries are odd or even. We can then summarize the effect of a Weyl reflection in the following prescription: The effect of $w_i$ is to add a multiple of the value of $p_i$ to the $p_k$ of the nodes adjacent (connected by lines) to $i$. If the line connecting the nodes $i$ and $k$ is single or triple, then $p_i$ has to be added to $p_k$. If the line is a double line, then $p_i$ has to be added to $p_k$ if $|\alpha_i|> |\alpha_k|$, but not otherwise. In graphical language: For double lines the value of $p_k$ can only go along double lines in the direction of the arrow.

Hence under Weyl reflections, each $p_k$ receives contributions from the adjacent nodes. Invariance under Weyl reflections implies choosing the $p_k$ such that all the contributions from adjacent nodes cancel out. An immediate consequence is for example that if the diagram has a chain with an endpoint (all the $\tp{G}$ Dynkin diagrams have at least one such endpoint), that the node adjacent to the endpoint must have $p_i=0$. For chains of single lines with endpoint this implies that every second node must have $p_i$. After these preliminary considerations it is easy to solve for the remaining $p_i$. 

We should note however that if $i(f)$ and $j(f)$ are indices in the above sense, that then also $(i+j)(f)$ is an index. Hence, when enumerating all the indices, we can restrict to a set of linearly independent indices.

For these indices we have not built in the invariance under outer automorphisms, and indeed, for the $\tp{D_n}$ algebra's there exist indices that are not invariant under these, and hence not under T-duality. We will exhibit specific indices for specific $\tp{G}$ algebra's later.

\section{$\tp{G}$ algebra's: explicit computations} \label{Gspecific}

We will now illustrate the general findings of the previous section, and fill in the details of their application to specific $\tp{G}$-algebra's. We will recover and extend the results of \cite{deBuyl:2005it}. First however we derive a simple result that will be highly useful later.

\subsection{An index for $SL(2n,\R)$ algebra's} \label{Lemma}

The following result turns out to be very useful. It is essentially formulated in terms of a "$\mod \ 2$" index for $SL(2n,\R)$ algebra; it is useful because in many instances one can identify a contribution of this sort to an index of a $\tp{G}$ algebra.

{\bf Lemma:} Consider the algebra $SL(2n,\R)$, which has as its Dynkin
diagram a straight line connecting $2n-1$ nodes. We number its nodes
$1,\ldots,2n-1$ from one end to the other. Consider an element $f \in
Q_{A_{2n-1}}$ and compute 
\be \label{I}
I = \inp{\sum_{i=1}^n \omega_{2i-1}}{f} \ \mod \ 2
\ee
The generators $e_{\alpha} - \exp ( i \pi \inp{\alpha}{f}e_{-\alpha}$ generate $SO(p, 2n-p)$ where $p$ is odd if $I=1$, and $p$ is even when $I = 0$.  

This can be proven quickly when we use an observation from \cite{Keurentjes:2004xx}. There it was noted, that for an $A_{D-1} = SL(D, \R)$ algebra, with a function $f$ written as
\be
f=\omega_{k_1} + \omega_{k_2} + \ldots + \omega_{k_n} \qquad k_1 > k_2 > \ldots > k_n, 
\ee
 $f$ defines the denominator subgroup $SO(k,D-k)$ with $k$ described by an alternating sum of the $k_i$
\be \label{k}
k = \sum_{i=1}^n (-)^{i+1} k_i
\ee

We note that when computing modulo 2, $p = 2n-p$. Computing $(\ref{k})$ modulo 2, one can immediately omit the signs $(-)^{i+1}$ and then also skip all the even $k_i$. But then it is easy to see that 
\be
(I = k =p) \ \mod \ 2
\ee

This lemma is useful because among the $\tp{G}$ algebra's there are
many instances where one can identify a contribution of the form $I$
in $i_0(f)$. If this contribution corresponds to a gravitational
subalgebra, then $I$ distinguishes between the space-time signature
$(even,even)$ and $(odd,odd)$. These space-time signatures admit
(anti-)self dual forms with $n$-form field strength when $n \ \mod \ 2= I$. More precisely if $(n+I) \ \mod \ 2 =0$, then the square of the Hodge-star $\df{}^2 = 1$; if $(n+I) \mod \ 2 =1$ then the square of the Hodge star $\df{}^2 = -1$. When there are $n$-forms transforming in some representation of an internal symmetry group, then these representations have to be real, respectively complex, otherwise they cannot obey a relation of the form
\be
\df{F_{(n)}} = \pm F_{(n)}
\ee
In relevant situations, this reality or complexity of representations is implied in the $i_0(f)$ condition. 

\subsection{Simply-laced split $\tp{G}$ algebra's}

We now demonstrate a number of explicit computations. We will first discuss the simply laced $\tp{G}$ algebra's. As the gravity sub-algebra consists entirely of long roots, which can never mix with short roots under diagram automorphisms and under Weyl reflections, these are to some extent the most representative for the T-duality pattern. Moreover, in many aspects the non-simply laced and non-split algebra's can be viewed as an extension of the simply laced algebra's. Correspondingly, the theories based on non-simply-laced and non-split algebra's are extension of theories based on simply laced algebra's, by including extra matter sectors.

\subsubsection{$\tp{A_n}$ algebra's: General Relativity in various dimensions}

\begin{figure}[ht]
\begin{center}
\includegraphics[width=8cm]{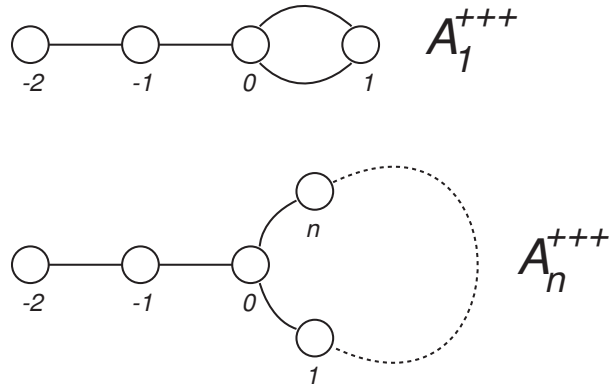}
\caption{Dynkin diagrams of the triple extended $A$-algebra's.}\label{Afig}
\end{center}
\end{figure}

The $\tp{A_n}$ algebra is conjectured to describe symmetries of General Relativity, formulated in $D=n+3$ dimensions.

The highest root of the $A_n$ algebra is given by 
\be
\alpha_H = \sum_{i=1}^n \alpha_i
\ee 
The consistency criterion for the index $i_0(f)$ hence takes the explicit form
\be
i_0(f) = \inp{\alpha_{-2}+ \alpha_0 + \sum_{i=1}^n \alpha_i}{f} \ \mod 2 = 1
\ee
The $\tp{A_n}$-algebra's with $n>1$ allow a diagram automorphism, whose action consists of exchanging $\alpha_i$ with $\alpha_{n+1-i}$ for $1 \leq i \leq n$. The index $i_0$ remains invariant under this automorphism.
  
Omitting the $\alpha_{n}$ node and all links to it leads to the Dynkin diagram of $SL(D)$, which is supposed to describe the simple part of the $GL(D)$ relevant to the vielbein. After embedding the space-time signature in this $SL(D)$-algebra, the $i_0$ condition fixes the remaining sign, so apart from the space-time signature there are no adjustable signs.

For $n$ is odd (General Relativity in even dimensions), there is an extra index:
\be
i_1(f) = \inp{\sum_{k=0}^{(n-1)/2} \alpha_{2k+1}}{f} \ \mod \ 2
\ee
Its meaning is easier described in terms of the index $i_0(f) + i_1(f)$. With the $\alpha_n$ node omitted, this index is of the form $I$ we have used in the lemma in subsection \ref{Lemma}. Hence we can link the values of $i_0+i_1$ to the space-time signature. However, consistency requires that $i_0(f) = 1$ always, and hence $i_0(f) + i_1(f) = i_1(f) +1$. So we have $i_1(f)= 1$ if the space-time signature $(t,s) = (even,even)$, and $i_1(f) = 0$ for $(t,s)=(odd, odd)$.

We can see immediately that 4-dimensional General Relativity, governed by $\tp{A_1}$ does \emph{not} allow
signature changing dualities. This is proven in two steps: one first notes that Weyl reflections in
$\alpha_i$ $(i=1,2,3)$ do not affect the space-time signature (as they only permute directions); the Weyl reflection in $\alpha_4$ can alter the $f$-function by integer multiples of $2 \alpha_4 \in 2 Q(\tp{A_1})$, but this does not affect the space-time signature, which is a $\mod \ 2$ computation. The impossibility to perform a time-like T-duality could also have been anticipated from our 
physical interpretation: For general relativity T-duality has to be generated by Kaluza-Klein-monopoles. But for 4-dimensional General Relativity these are not extended objects, so there cannot be any relation via "string-dualities".

For higher dimensional General Relativity, we can immediately check time-like dualities by embedding the space-time signature in an appropriate way.

We start with considering the signature $(0,D)$. The requirement $i_0(f)=1$ implies that the right real form of the $K(\tp{A_n})$ algebra must be a non-compact form given by:
\be
f=\omega_{n}
\ee
Weyl reflecting in $\alpha_n$ and reducing modulo 2 results in 
\be
f = \omega_0 + \omega_{n-1}+ \omega_n.
\ee
This is easily seen to represent signature $(D-4,4)$, and hence reproduces the naive argument of section \ref{dual}. It is amusing to note that this computation even applies for $D=4$; the formula's are predicting that this theory does not transform under timelike T-duality. 
The signature $(p,D-p)$ can be implemented by (if $1 <p \leq n-1 = D-4$)
\be
f=\omega_0+ \omega_{p} + \omega_n
\ee
Weyl reflecting in $\alpha_n$ leads to
\be
f = \omega_{p} + \omega_{n-1} + \omega_n
\ee
This represents signature $(D-4-p,p+4)$, which agrees with the derivation in section \ref{dual}. Note that this duality becomes impossible if $p > D-4$.

If $2 \leq p \leq D-2 = n+1$ we can also embed the $(p,D-p)$ by using
\be
f = \omega_{-2} + \omega_{p-2} + \omega_n
\ee
Now the Weyl reflection in $\alpha_n$ leads to 
\be
f = \omega_{-2} + \omega_{0} + \omega_{p-2} + \omega_{n-1} + \omega_n
\ee
which represents space-time signature $(D-p,p)$.

Collecting all the possible orbits under time-like T-duality, we obtain the following table, which is valid for $D \geq 7$:
\be
\begin{tabular}{|c|c|}
\hline
$D \  \mod \ 4$  & $(t \ \mod \ 4,s \ \mod \ 4)$ \\
\hline
\hline
$0$ & $(0, 0)$ \\
     & $(1, 3) \leftrightarrow (3,1)$ \\
     & $(2,2)$\\
\hline
$1$ & $(0,1) \leftrightarrow (1,0)$ \\
    & $(2,3) \leftrightarrow (3,2)$ \\
\hline
$2$ & $(0,2) \leftrightarrow (2,0)$ \\
    & $(1,1)$ \\
    & $(3,3)$ \\
\hline
$3$ & $(0,3) \leftrightarrow (3,0)$ \\
    & $(1,2) \leftrightarrow (2,1)$ \\
\hline
\end{tabular}
\ee
We have required $D \geq 7$ because for $D=4,5,6$ some of these orbits break up due to the absence of the relevant duality operation. For $D=4$ timelike T-dualities are impossible, as explained previously. For the dimensions $D=5,6$ the orbits are
\be
\begin{tabular}{|c|c|}
\hline
$D$  & $(t,s)$ \\
\hline
\hline
$5$ & $(0, 5) \leftrightarrow (1,4)$ \\
     & $(2, 3) \leftrightarrow (3,2)$ \\
     & $(4,1) \leftrightarrow (5,0)$\\
\hline
$6$ & $(0,6) \leftrightarrow (4,2) \leftrightarrow (2,4) \leftrightarrow (6,0)$ \\
    & $(1,5)$ \\
    & $(3,3)$ \\
    & $(5,1)$ \\
\hline
\end{tabular}
\ee

These orbits (some of which can also be found in \cite{deBuyl:2005it}) of the gravitational algebra are of importance because any $\tp{G}$ algebra has a $\tp{A_n}$ algebra as subalgebra. For \emph{any} of these theories the signatures listed here should be connected by duality transformations. The only issue may be whether some of these orbits may get connected due to additional symmetries.

\subsubsection{$\tp{D_n}$-algebra's: String theories in various dimensions}

\begin{figure}[ht]
\begin{center}
\includegraphics[width=12cm]{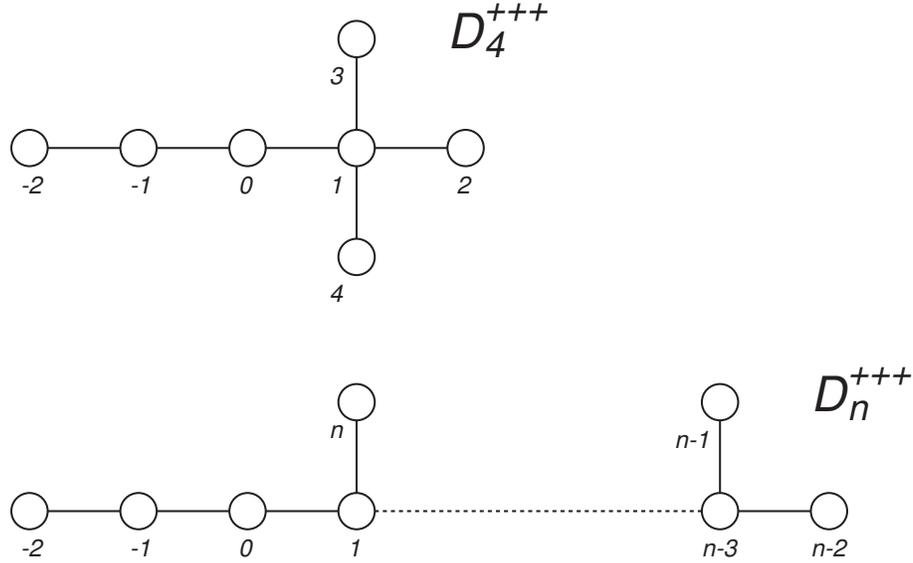}
\caption{Dynkin diagrams of the triple extended $D$-algebra's.}\label{Dfig}
\end{center}
\end{figure}

The $\tp{D_n}$ algebra's (with $n \geq 4$) are conjectured to represent bosonic string theory in $D=n+2$ dimensions. 

The highest root of a $D_n$ algebra reads
\be
\alpha_H = \left(\sum_{i=1}^{n-3} 2 \alpha_i \right) + \alpha_{n-2} + \alpha_{n-1} + \alpha_n
\ee 
The $i_0$ criterion reads:
\be
i_0(f) = \inp{\alpha_{-2} + \alpha_0 + \alpha_{n-2} + \alpha_{n-1} + \alpha_n}{f} \ \mod \ 2 =1
\ee
The $\tp{D_n}$ algebra's always have multiple indices. There is always the second index
\be
i_1(f) = \inp{\alpha_{n-2} + \alpha_{n-1}}{f} \ \mod \ 2
\ee
Moreover for $n > 4$ and even there is a third independent index
\be
i_2(f) = \inp{\sum_{k=1}^{n/2-2} \alpha_{2k} + \alpha_{n-1} + \alpha_n}{f} \ \mod \ 2
\ee
The $\tp{D_n}$ algebra's with $n >4$ allow a diagram automorphism that is inherited from the $D_n$ algebra. It exchanges the nodes $\alpha_{n-1}$ and $\alpha_{n-2}$. Hence it leaves $i_0$ and $i_1$ invariant, but it exchanges $i_2$ with $i_1 + i_2$. 

The $\tp{D_4}$ algebra has a highly symmetric diagram, and has apart from $i_0(f)$ the indices
\be
i_1(f) = \inp{\alpha_2 + \alpha_3}{f} \qquad i_2(f) = \inp {\alpha_2 + \alpha_4}{f} \ \mod \ 2
\ee
The algebra $\tp{D_4}$ allows an even bigger set of automorphisms,
inherited from the triality of $D_4$: Any permutation of $\alpha_2$,
$\alpha_3$ and $\alpha_4$ corresponds to an admissible
automorphism. All of these leave $i_0$ invariant, but they permute
$i_1$, $i_2$ and $i_1+i_2$.

In the theory with maximal dimension $D=n+2$, the $SL(D)$ chain
representing gravity is formed by the nodes $\alpha_{-2}, \ldots ,
\alpha_{n-2}$. It is also possible to form a $D$-dimensional theory by
instead using the chain $\alpha_{-2}, \ldots, \alpha_{n-3},
\alpha_{n-1}$. This formulation is related to the previous one by an
outer automorphism of $\tp{D_n}$, or equivalently, the two are related by a T-duality.
Whichever way we choose to embed the chain, there are two nodes not
included. For concreteness, we choose these to be $\alpha_{n-1}$ and
$\alpha_n$. The $i_0(f)$ condition fixes the product of their signs,
but we can choose one sign independently. Physically, these nodes
correspond to a 2-form in the maximal dimension ($\alpha_{n-1}$), and
a $D-4$-form ($\alpha_n$), that is the Poincar\'e dual to the
2-form. Poincar\'e duality tells us their signs should be related, and
indeed, so does its algebraic counterpart, the $i_0$ condition. The
index $i_1(f)$ incorporates the sign of the 2-form, but not of the
$(D-4)$-form, and hence is an interesting quantity if we wish to keep
track of the signs. If we specify the space-time signature and
$i_1(f)$ the signs of the theory are essentially fixed. We note also
that the roots corresponding to the "dual of the graviton" are at
higher levels, but that essentially its sign is the product of the
signs of the 2-form with the sign of the $(D-4)$-form.

In Euclidean signature (0,D) we have the two possibilities 
\be
f = \omega_{n-1} \ \leftrightarrow \ i_1(f)=1 \qquad \textrm{or} \qquad f = \omega_{n} \ \leftrightarrow \ i_1(f)=0
\ee
The option $f=\omega_{n-1}$ allows a Weyl-reflection in the $\alpha_{n-1}$ (string-)node, resulting in the signature $(2,D-2)$. Weyl reflections in the node $\alpha_{n}$ (the $(D-5)$-brane node) have no signature changing effect. Note that with these conventions the $2$-form kinetic term has a minus sign, and hence allows extremal string solutions of world volume signature $(0,2)$. That implies however that the $(D-4)$-form field term has the conventional sign, so a $(D-5)$ brane must have an odd number of time directions on its world volume, which is impossible.

For the choice $f= \omega_n$ the roles are reversed: Now the kinetic term for the 2-form has a plus-sign and the $(D-4)$-form a minus-sign, implying the existence of a $(D-5)$-brane but absence of a string. Now a signature changing duality induced by a Weyl reflection in the node $\alpha_n$ is possible, resulting in the signature $(D-4,4)$.

We now turn to the signatures $(p,D-p)$ $1 \leq p \leq D-3=n-4$, and with $i_1(f) = 0$. We set
\be
f  =  \omega_{n-3} + \omega_{n-2} + \omega_{n-1}, 
\ee
for $p=1$, and
\be
f  =  \omega_{n-2-p} + \omega_{n-3} + \omega_{n-2} + \omega_{n-1} + q \omega_n, 
\ee
for $1 < p \leq n-4 = D-3$, where $q$ can be adjusted to make $i_0(f)=1$. Weyl reflecting in the string node $\alpha_{n-1}$ amounts to the disappearing of $\omega_{n-3}$ from the above expressions. This however does not have any effect on the signature. This is because the relevant theories all have a conventional sign in front of the $2$-form, implying a string with world-volume of signature $(1,1)$. The rules we outlined in section \ref{dual} then imply that the overall space-time signature cannot change under reflections in these nodes. Note also that $i_1(f) = 0$ implies the same for the T-dual theory, the one where we interchange the roles of $\alpha_{n-1}$ and $\alpha_{n-2}$.

The theory with $p=1$ has the standard Minkowski signature, and here we can generate time-like T-dualities only with the 5-brane node (and hence time-like T-duality must be invisible in CFT \cite{Moore:1993zc}). To study duality in the 5-brane node we set ($1 \leq p < n-4 = D-6$)
\be
f= \omega_1 + \omega_{p+1} + \omega_n
\ee
These theories also have signature $(p,D-p)$ and $i_1(f) =1$, and it is not hard to show that they are connected to the previous theories by permutation of the coordinates (Weyl reflections in nodes other than $\alpha_{n-1}$ and $\alpha_n$). A Weyl reflection in $\alpha_n$ will result in
\be 
f = \omega_{p+1} + \omega_n
\ee
representing signature $(D-p-4,p+4)$. This is generated by a reflection corresponding to a $(D-4)$-brane with world volume signature $(p,D-p-4)$:
\be
(p,D-p) = (p,D-p-4) + (0,4) \rightarrow (D-p-4,p) + (0,4) = (D-p-4,p+4)
\ee
The $(D-4)$-brane solutions with these world volume signatures ought to exist provided the sign in front of the corresponding kinetic term is $(-)^{p+1}$, which of course perfectly correlates with our claim that the 2-form terms had a plus-sign for their kinetic terms. 

The other subcases correspond to $i_1(f) = 1$. To study duality in the node $(n-1)$ we set 
\be
f= \omega_{n-3-p} + \omega_{n-3} + \omega_{n-1} + q \omega_n
\ee
where again $q$ should be set to 0 or 1 to fulfill the $i_0$-condition. These represent space-time signatures $(p, D-p)$ with $1 \leq p \leq n-1 = D-3$. A Weyl reflection in the node $\alpha_{n-1}$ leads to
\be
f= \omega_{n-3-p} + \omega_{n-1} + q \omega_n
\ee
which represents the space-time signature $(p+2, D-p+2)$. This is consistent with the interpretation of the theory as one with wrong signed 2-form term, which implies string solutions with world-volume signature $(2,0)$ and $(0,2)$.

To study duality in the $(D-5)$-brane node when $i_1(f) = 1$ we set
\be
f= \omega_0 + \omega_p + \omega_{n-1} + \omega_n
\ee
with $1 \leq p \leq n-3 = D-5$. This represents signature $(p,D-p)$. Reflecting in $\alpha_n$ leads to
\be
f= \omega_0 + \omega_1 + \omega_{p} + \omega_{n-1} + \omega_n
\ee
which represents the signature $(D-p-2,p+2)$. This implies there is a $(D-5)$-brane with world volume signature $(p-1,D-p-3)$, meaning that the sign for the kinetic terms of the $(D-6)$-form is $(-)^{p}$, perfectly consistent with our claim that the theories with $i_1(f)$ have a wrong signed kinetic term for the 2-form.

There are many other signature changes possible, that we will not exhibit explicitly. 

The theories conjectured to be described by $\tp{D_n}$ algebra's allow another interesting oxidation branch, leading to 6 dimensional theories (see \cite{Keurentjes:2002xc}.  The gravitational subalgebra is now the one corresponding to the nodes $\alpha_{-2}, \alpha_{-1}, \alpha_0, \alpha_1$ and $\alpha_n$. The space-time signature is then governed by $i_0(f)+i_1(f)$, if this is 1 then the signature is $(odd,odd)$, if it is 0 then the signature is $(even,even)$. With the constraint $i_0(f)=1$, this implies that the space-time signature fixes $i_1(f)$. The reason for this is that the theory admits 3-form field strengths, that should be subjected to a (anti-)self-duality constraint. For 6 dimensional space-time signature $(even,even)$ the eigenvalues of the Hodge star are imaginary, and $i_1(f) = 1$. One can verify, using the techniques of \cite{Keurentjes:2004xx} (see also the appendix of this paper), that this implies that the 6 dimensional coset symmetry is $SO(n-3,n-3)/SO(n-3,\C)$, such that the representation of the 3-forms are indeed complex valued. In contrast, when $i_1(f) = 0$, and the space-time signature is $(odd,odd)$, the coset is of the form $SO(n-3,n-3)/SO(n-3-p,p) \times SO(n-3-p,p)$, and the representations to which the 3-forms belong is real.

\subsubsection{$\tp{E_6}$: Membranes in 8 dimensions}

\begin{figure}[ht]
\begin{center}
\includegraphics[width=14cm]{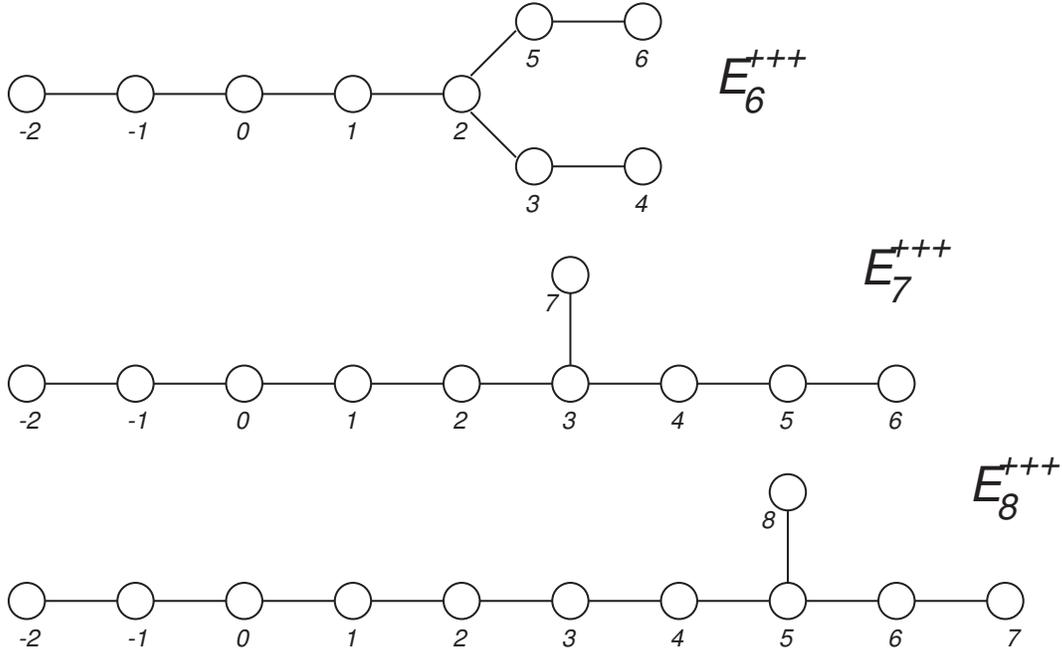}
\caption{Dynkin diagrams of the triple extended $E$-algebra's.}\label{Efig}
\end{center}
\end{figure}

To the $\tp{E_6}$-algebra corresponds an 8 dimensional theory that has a doublet of 3-forms. This doublet transforms under $SL(2)/SO(2)$ or $SL(2)/SO(1,1)$, depending on the choice of various signs.

The highest root of the $E_6$ algebra is
\be
\alpha_H = 2 \alpha_1 + 3 \alpha_2 + 2 \alpha_3 + \alpha_4 + 2\alpha_5 + \alpha_6
\ee
The $\tp{E_6}$-algebra only has one index
\be
i_0(f) = \inp{\alpha_{-2} + \alpha_0 + \alpha_{2} + \alpha_4 + \alpha_6}{f} \  \mod \ 2 =1
\ee
The $i_0$-condition then tells us that if the signature of the 8-dimensional space-time is $(odd,odd)$, that the coset must be of the form $SL(2),\R)/SO(2)$, while when the signature is $(even,even)$ the coset has to be $SL(2)/SO(1,1)$. This correlates perfectly with Poincar\'e duality: In signature $(odd,odd)$ the eigenvalues of the duality operation are $\pm i$, and the 3-form and its dual combine in a complex combination, rotated into one another by $SO(2) \cong U(1)$. In signature $(even,even)$ however the eigenvalues of Hodge-star are $\pm 1$, which is consistent with the fact that the $SO(1,1)$ algebra is real.

Note that according to our rules, a doublet of 3-forms in 8 dimensions gives rise to a doublet of membranes, which gives rise to a dual theory with a doublet of membranes. Of course the dual theory is related to the original one by the outer automorphism of $\tp{E_6}$. The analysis of the dual theory is identical to that for the original theory, due to the fact that $i_0(f)$ is invariant.

Once we have chosen a space-time signature, the $i_0$ condition tells us whether the 8-dimensional coset is $SL(2,\R)/SO(2)$ or $SL(2,\R)/SO(1,1)$. If it is $SL(2,\R)/SO(2)$, we still have some freedom in choosing consistent signs. The 8-dimensional theory has a doublet of 3-forms (coupling to membranes), and the kinetic term for this doublet can have either a plus or minus sign. This sign is encoded in a contribution of $\omega_5$ to $f$. In the case the coset is $SL(2,\R)/SO(1,1)$ the doublet has 1 plus and 1 minus sign, and the relative sign can be changed by a $SL(2,\R)/SO(1,1)$ transformation, so there are no additional signs to be adjusted. 

A detailed analysis of the possible signature and how they connect under duality transformations/Weyl reflections gives the following two orbits:
\be \label{Houart}
(1,7,+), \quad (2,6), \quad (3,5,+), \quad (5,3,+), \quad (6,2), \quad (7,1,+).
\ee
\be
(0,8), \quad (1,7,-), \quad (3,5,-), \quad (4,4), \quad (5,3,-), \quad (7,1,-), \quad (8,0);
\ee
Because only in the case where the space-time signature is $(odd, odd)$ there are unspecified signs, we have added to these signatures a $+$ or $-$ indicating whether the 3-form kinetic terms come with conventional ($+$) or unconventional sign. The first of these orbits is also discussed in \cite{deBuyl:2005it}, where another notation for the signs is used.
 
\subsubsection{$\tp{E_7}$: 3-branes in 10-dimensions; or strings in 8-dimensions}

The highest root of the $E_7$ algebra is given by
\be
\alpha_H = 2 \alpha_1 + 3 \alpha_2 + 4 \alpha_3 + 3\alpha_4 + 2 \alpha_5+ \alpha_6 + 2 \alpha_7
\ee
The $i_0(f)$ index for $\tp{E_7}$ then reads
\be
i_0(f) = \inp{\alpha_{-2} + \alpha_0 + \alpha_2 + \alpha_4 + \alpha_6}{f} \ \mod \ 2= 1
\ee
A second index is given by
\be
i_1(f) = \inp{\alpha_4 + \alpha_6 + \alpha_7}{f} \ \mod \ 2. 
\ee

The $\tp{E_7}$-algebra leads to a 10-dimensional theory with a self-dual 4-form\footnote{In many references this theory is not oxidized beyond 9 dimensions, but the existence of the 10-dimensional version is well-known \cite{Cremmer:1999du, Keurentjes:2002xc}. A simple way to construct the theory is by starting with the bosonic sector of type IIB theory, and then omitting the scalar sector and truncating further to the fields that are singlets under $SO(2)/SO(1,1)$.}. Self-duality is only consistent with signature $(odd,odd)$, but then any sign for the kinetic terms for the 3-form is possible. This is indeed exactly the content of the $i_0$-condition for this theory. Once we fix the space-time signature, the sign of the 4-form kinetic term is encoded in $i_1$. From this we can learn that once the sign is fixed to plus or minus, there does not exist a time-like T-duality to a 10 dimensional theory where the 4-form kinetic term has opposite sign.

The 3-brane coupling to the 4-form in 10 dimensions, dualizes to a
 triplet of strings coupling to 2-forms in 8 dimensions. Indeed, by
 choosing the $SL(8)$-chain running from the $\alpha_{-2}$ node to the
 $\alpha_7$ node, we find this second, dual theory. The indices
 for these theory are easier discussed if we first look at the index $i_0 + i_1$. It is easy to see that if $i_0 + i_1=1$ (which, because $i_0 = 1$ corresponds to $i_1 = 0$), this corresponds to space-time signature $(odd,odd)$ for the 8-dimensional theory, while $i_0+ i_1 =0$ ($i_1 = 1$) corresponds to space-time signature $(even, even)$.

The 10-dimensional theories are connected in the following way under time-like T-dualities. We indicate a $-$ if the value of $i_1=1$ and a $+$ if $i_1=0$. Then the orbits are
\be
(3,7,-) \leftrightarrow (7,3,-)
\ee
\be
(1,9,-) \leftrightarrow (5,5,-) \leftrightarrow (9,1,-)
\ee
All of these theories have 8-dimensional duals that have space-time signature $(even,even)$. 

The more conventional theories have $i_1(f)=0$, and can all be dualized into each other, that is the signatures 
\be
(1,9,+), \quad (3,7,+), \quad (5,5,+), \quad (7,3,+), \quad (9,1,+),
\ee
are all connected. Because of $i_1(f)=0$, these must dualize to 8-dimensional theories with space-time signature $(odd,odd)$.

\subsubsection{$\tp{E_8}$: M-theory and IIB-string theory}

In \cite{Keurentjes:2004bv} the possible signatures and orbits for the M-theory interpretation of the $\tp{E_8}$ algebra's where listed. These remain correct, but we are arguing in the present paper that there is an extra consistency condition. As the highest root for the $E_8$ algebra is
\be
\alpha_H = 2 \alpha_1 + 3 \alpha_2 + 4 \alpha_3 + 5 \alpha_4 + 6 \alpha_5 + 4 \alpha_6 + 2 \alpha_7 + 3 \alpha_8
\ee
this extra condition, takes the following form for $\tp{E_8}$
\be
i_0(f) = \inp{\alpha_{-2} + \alpha_0 + \alpha_2 + \alpha_4 + \alpha_8}{f} \ \mod \ 2= 1
\ee
These implies that some of the generalized signatures in \cite{Keurentjes:2004bv} are inconsistent with Poincar\'e duality. The remaining ones organize in the orbits with signatures (note that the $i_0(f)$- condition fixes the sign of the 3-form to $(-)^{T-1}$ where $T$ is the number of time directions!)
\be
(1,10); \quad (2,9); \quad (5,6); \quad (6,5); \quad (9,2); \quad (10,1)
\ee
and 
\be
(0,11); \quad (3,8); \quad (4,7); \quad (7,4); \quad (8,3); \quad (11,0).
\ee

The first orbit contains the standard signatures of the $M$, $M'$ and $M^*$ theories \cite{Hull:1998ym}. 

The second orbit contains precisely the remaining signatures. Though it is standard lore that these 11-dimensional theories are non-supersymmetric, some of the 10 dimensional theories included here seem to have supersymmetric extensions \cite{Vaula:2002cn}.

The various signatures for the IIB theories can be derived straightforwardly. Note that the $i_0(f)$ condition states that the space-time signature for a IIB-theory or a variant must be $(odd, odd)$. Variants of the IIB-theory cannot occur in signatures $(even,even)$ because the 4-form whose field strength will be (anti-)self dual is a singlet under the $SL(2)$ coset symmetries; we cannot tune its duality properties. The orbit with M-theories contains all the signatures and sign patterns of the $IIB$, $IIB'$ and $IIB^*$ theories \cite{Hull:1998vg, Hull:1998ym}, the other orbit contains the signatures $(1,9,-,-)$, $(3,7,+,+)$, $(3,7,+,-)$, $(5,5,+,+)$, $(7,3,+,+)$, $(7,3,+,-)$, and $(9,1,-,-)$ (in the notation of \cite{Keurentjes:2004xx}). 

\subsection{Non-simply laced and non-split $\tp{G}$-algebra's}

\subsubsection{Non-simply laced split}

\begin{figure}[ht]
\begin{center}
\includegraphics[width=12cm]{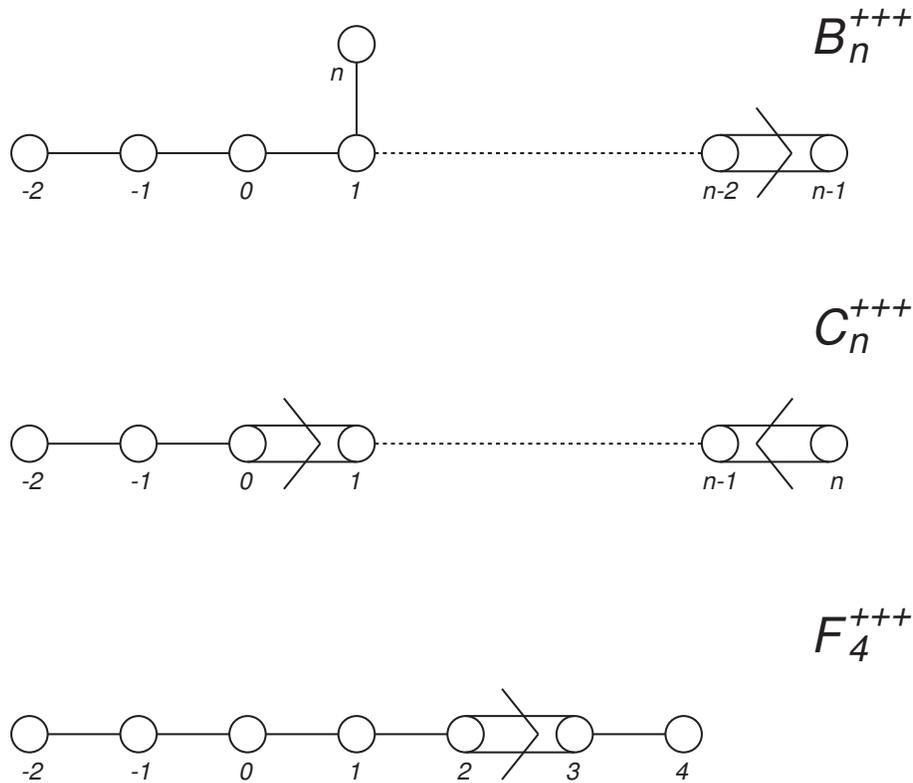}
\caption{Dynkin diagrams for the triple extended $2$-laced algebra's.}\label{2fig}
\end{center}
\end{figure}

The non-simply laced algebra's are distinguished by the appearance of short roots. Short roots can never be exchanged with the gravity sector (as these correspond to long roots). For the space time signature these are hence of no interest. Signs for short roots are possible, but these just represent wrong signs in the matter sector. The time-like T-duality pattern follows in principle from the long roots. Of importance is therefore the largest simply laced algebra inside the non-simply laced algebra's. These are given by
\be
\tp{B_n} \supset \tp{D_n} \qquad \tp{F_4} \supset \tp{D_4}
\ee
The long roots of $\tp{C_n}$ are in some sense given by a "triple
extension of $A_1^n$". This algebra however does not allow a Dynkin
diagram\footnote{The same is true for any non-simple group, which is
  probably one of the reasons why these theories have not been
  considered much so far. Note that it is completely straightforward
  to define theories with non-simple groups in 3-dimensions; it is
  their "triple extension" that is hard to define.} (curious readers
may try $n=2$), and the easiest way to describe the corresponding
algebra seems to be as a truncation of the $\tp{C_n}$ algebra, so this
does not have many advantages.  

As the highest root for the $B_n$ algebra is
\be
\alpha_H = \left(\sum_{i=1}^{n-1} 2 \alpha_i \right) + \alpha_n,
\ee 
the index $i_0(f)$ for the $\tp{B_n}$ algebra is given by
\be
i_0(f) = \inp{\alpha_{-2} + \alpha_{0} + \alpha_n}{f} \ \mod \ 2=1
\ee
A second index is given by 
\be
i_1(f) = \inp{\alpha_n}{f} \ \mod \ 2.
\ee
This is invariant due to the 2-laced nature of the algebra.

The theory allows two maximally oxidized variants. There is a
$D=n+2$-dimensional theory, which contains General Relativity, a
vector and its dual $D-3$-form, and a 2-form and its dual
$(D-4)$-form. The sign of the kinetic term for the 2-form cannot be
chosen freely; the algebra requires the sign to be the square of the
sign for the kinetic term of the vector, and hence this is always
$+$. Because of this, the $i_0(f)$-condition fixes the sign of the
$(D-3)$-form once the space-time signature is embedded in the chain
from $\alpha_{-2}$ to $\alpha_{n-2}$. None of these considerations
fixes the sign of the vector, this is encoded in a separate index
$i_1(f)$. 

These theories can easily be seen, both from the algebraic as from the
physical perspective, to have the same time-like T-duality pattern as
the theories with $\tp{D_n}$ algebra, and $i_1(f)=1$ (because the
two-form sign is fixed!), so we refer the reader to the discussion
there. 

A second branch for this theory gives a theory in 6 dimensions, with
$(2n-5)$ 2-forms satisfying (anti-)self duality constraints. These
transform in the vector representation of a real form of
$B_{n-3}$. The $i_0(f)$-condition implies that the signature of the
6-dimensional theory is $(odd,odd)$. The $(even,even)$ signatures
cannot be realized; these would require complex representations for
the 2-forms, but the relevant representations (the vector
representations $\mathbf{(n-2,1) \oplus (1,n-3)}$ of some real form of
$SO(n-2) \times SO(n-3)$) are always real regardless of the choice of
$f$.

For the $C_n$ algebra's the highest root is
\be
\alpha_H= \left(\sum_{i=1}^{n-1}2 \alpha_i \right) + \alpha_n
\ee
For the $\tp{C_n}$ algebra the index $i_0(f)$ is given by
\be
i_0(f) = \inp{\alpha_{-2} + \alpha_{0} + \alpha_n}{f} \ mod \ 2=1
\ee
For $n$ is even a second index reads
\be
i_1(f) = \inp{\sum_{i=1}^{n/2} \alpha_{2i-1}}{f} \ \mod  \ 2.
\ee
Associated to these algebra's are 4 dimensional theories of General
Relativity, coupled to a sigma model on $Sp(n-1,\R)/H$ and a bunch of
vectors and their duals transforming in the fundamental
$2n$-dimensional  representation. The denominator subgroup is a real
form of $GL(n-1,\C)$. It can be $GL(n,\R)$, or $U(n-p,p)$ for some
value $p$ (see appendix A of \cite{Keurentjes:2004xx}). We immediately
note that $GL(n,\R)$ is a group over the reals, while the
representations of $U(n-p,p)$ are typically complex. The $i_0(f)$
condition relates the possible real form of $GL(n, \C)$ to the
space-time signature. To be precise, for space-time signature
$(even,even)$ the denominator subgroup in 4 dimensions must be a real
group (because of duality requirements), and it can be shown that the
$i_0(f)$ condition implies it is $GL(n,\R)$. For space-time signature
$(odd,odd)$ however, the $i_0(f)$-condition, as well as duality
requirements state that $H$ must be complex (and hence, of the form
$U(n-p,p)$. 

The index $i_1$ is correlated to the fact that we can change the sign of the vectors in 4 dimensions; only for $\tp{C_n}$ theories with $n$ even this leads to a conserved quantity.

The time-like T-duality pattern is rather boring, because it is the
same as for $\tp{A_1}$ (4-dimensional general relativity) that does
not allow any time-like T-dualities.

For $F_4$ the highest root is
\be
\alpha_H = 2\alpha_1 + 3 \alpha_2 + 4 \alpha_3 + 2 \alpha_4
\ee
For $\tp{F_4}$ the index $i_0(f)$ is given by
\be
i_0(f) = \inp{\alpha_{-2} + \alpha_{0} + \alpha_2}{f} \ \mod \ 2=1
\ee
This algebra gives rise to a 6-dimensional theory with $SL(2,\R)$ symmetry, with vectors (and their dual 3-forms) transforming in the $\mathbf{2}$ and 2-tensors in the $\mathbf{3}$. The index $i_0(f)$ implies that the 6-dimensional theory has signature $(odd,odd)$. The $(even,even)$ signatures are again impossible, because the sign of the 2-forms is again the square of the vectors hence $+$. It is easy to establish that it is impossible to change the space-time signature. As we have argued the time-like T-duality pattern is the same as for $\tp{D}_4$; we however have to take into account the extra restriction that the space-time signature is fixed to $(odd,odd)$, which renders any signature change impossible. 

Note that signs from the gravity and fields described by long roots can run into the matter sector described by the short roots, which does allow sign changes in the matter sector. It is the opposite direction that is impossible.

\begin{figure}[ht]
\begin{center}
\includegraphics[width=12cm]{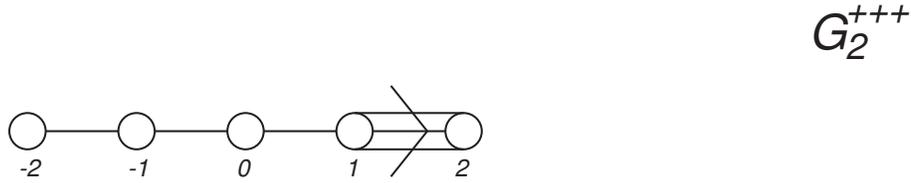}
\caption{Dynkin diagram of the triple extended $3$-laced algebra's.}\label{3fig}
\end{center}
\end{figure}
There is only one 3-laced algebra among the triple extended algebra's, which is $\tp{G_2}$. The highest root of $G_2$ is 
\be
\alpha_H = 2 \alpha_1 + 3 \alpha_2
\ee
The index condition is
\be
i_0(f) = \inp{\alpha_{-2} + \alpha_0 + \alpha_2}{f} \ \mod \ 2 =1
\ee
The oxidation endpoint of this theory is 5 dimensional Einstein-Maxwell gravity. The $i_0$ condition implies that the sign of the vector in 5 dimensions is determined by the space-time signature. This is because the dual 2-form is at level 2, and hence its sign cannot be adjusted. The $i_0$ condition states 
essentially that the sign of the vector terms is in accordance to the (fixed) sign of its dual. With this proviso, every space-time signature is possible. A computation shows that the orbits under duality are essentially the same as those for 5 dimensional gravity.

\subsubsection{Non-split groups}

One advantage of the derivation of the criterion $i_0(f)$, is that it
is clear that it immediately generalizes to theories that are not
based on split real forms of the algebra $G$ (for a survey that
encompasses all the simple algebra's that are not split, and the
corresponding theories see \cite{Keurentjes:2002rc}). We can allow $G$ to
be any real form as long as it is non-compact. One starts from coset
sigma model in 3 dimensions on $G/H$ (we will specify the possible $H$
in a moment), coupled to 3-dimensional general relativity. The obvious
conjecture is that such theories fit into the framework of $\tp{G}$
algebra's by extending their "Tits-Satake diagram" (see
e.g. \cite{Helgason}) by three more nodes. A Tits-Satake diagram is
nothing but a Dynkin diagram with decoration, so we extend the Tits-
Satake diagram as we would for the Dynkin diagram, while leaving the
extra nodes without decoration\footnote{It is conceivable that the
  extra nodes could be decorated. But then an extension of the
  analysis of \cite{Keurentjes:2002rc} combined with the usual
  framework of $\tp{G}$ algebra's (extraction of matter content via
  level decomposition, etc.) would lead us to conclude that these
  theories cannot be oxidized to theories in dimension 3 or
  higher. This hypothetical class of theories would hence be poorly
  understood both from the mathematical \emph{and} the physical
  side....} (in \cite{Keurentjes:2002rc} we already defined (once)
extended Tits-Satake diagrams).

Also here we could specify the signature in a function $f$, in the same way as we have done in \cite{Keurentjes:2004bv}. The only thing we wish to avoid is that $f$ takes non-trivial values on the invariant subspace of the Cartan-involution. If we, as in \cite{Keurentjes:2004xx} inscribe the values of $f$ on the Tits-Satake diagram, it is clear that we should only allow $f$ to have non-trivial values at white nodes, that are not decorated by arrows.

As long as there exists a 3-dimensional theory, our derivation of section \ref{Ggeneral} remains valid; we never used any information about the real form of $G$ there. So also in this case we should have $i_0(f)=1$.
We will not attempt any detailed analysis for this (vast) category of theories.

\section{Discussion and conclusions}\label{Con}

In this final section we describe, before reaching our conclusions, a number of issues that should provide the reader with food for thought, and possibly directions for future research. 

\subsection{The meaning of the $i_0(f)$-condition}

We have derived and presented the $i_0(f)$ condition as a necessary
condition to implement the action of Poincar\'e duality on $\tp{G}$
algebra's in the correct way. Though of course the relation to
Poincar\'e duality is sufficient to highlight the importance of
$i_0(f)$, there are further questions that can and should be posed.

One observation is that the requirement of Poincar\'e duality is in a
sense "external"; we have presented it as an argument related to the
interpretation of the algebra as that of a physical theory. There it
is rooted in the calculus of differential forms, which does not seem
to lead to a direct relation with the $\tp{G}$ algebra. Also the fact that we have derived it for a
particular decomposition, related to 3-dimensional theories and then
relied on physical arguments to extrapolate to all theories that can
be reconstructed from $\tp{G}$ algebra's is somewhat dissatisfying.
One would like to have a more "internal" explanation, where the
condition would follow as a consistency requirement from some
framework to set up the theory directly from the algebra. The only
such framework with manifest $\tp{G}$-symmetry that has been proposed
thus far \cite{Englert:2003py} does not really seem to require the
$i_0(f)$ condition from the outset, and one may (and probably should) wonder whether the
condition is hidden somewhere at a deeper level of the interpretation.

Another interesting issue is what the precise mathematical meaning of
the $i_0(f)$ condition is: What exactly is the distinguishing feature
that separates $K(\tp{G})$ defined by an $f$ with $i_0(f)=1$ from
those with an $i_0(f)=0$? In the application to theories with
(anti-)self-dual forms, the $i_0(f)$ condition ensures that the
eigenvalues of the Hodge-star, and in particular whether they are real
or imaginary, can be properly represented on the $K(\tp{G})$
algebra. This leads us to suspect that the $i_0(f)$ condition may be
some sort of "reality condition" on $K(\tp{G})$. Whether this is the
case could be verified if we knew the representation theory for
$K(\tp{G})$, but at present too little is known about these
algebra's.  

The cosets $\tp{G}/K(\tp{G})$ hypothetically describe a theory which
originally had a finite number of degrees of freedom, by an infinite number of
variables. It is an open question whether this should be interpreted
as an indication for new degrees of freedom, or that the finite number
of degrees of freedom should arise after imposing (infinitely many ?)
constraints (as in proposals that suggest that the higher level fields
contain the derivatives of lower level fields \cite{Damour:2002cu,
  Nicolai:2003fw, Englert:2003py}). It is also possible that we may
both need constraints and new degrees of freedom. Most interpretations
agree at least on the fact that both forms and dual forms are
independently represented by $\tp{G}$ algebra's. This indicates that the
$i_0(f)$-condition either is one of the constraints necessary to
reduce the number of degrees of freedom, or is implied
by such constraints. Therefore we expect that it should play an
essential role in the formulation of the theory with manifest
$\tp{G}/K(\tp{G})$ symmetry.   

Yet another interesting issue that should be related to the
$i_0(f)$-condition is the ability to couple the theories to
spinors. It was observed in \cite{Keurentjes:2003yu} that these should
transform in representation of $K(\tp{G})$. This has not been made
very precise thus far (but see the constructions of
\cite{Nicolai:2004nv, deBuyl:2005zy} relating to infinite-dimensional
sub-algebra's of $K(\tp{G})$, and the attempt of \cite{Miemiec:2004iv}
to implement supersymmetry in some $\tp{G}$-algebra's). Here we note
the close relation of Poincar\'e duality to the properties of the
Levi-Civita anti-symmetric tensor, and its important consequences for
the representation theory of spinors, in particular their chirality properties. We
therefore suspect that the $i_0(f)$ condition is relevant in this
context. If the $i_0(f)$ condition is indeed some sort of reality
condition on $K(\tp{G})$, as we have conjectured, then
also the existence of (pseudo-)Majorana spinors in the low-energy
effective theory may be encoded by it. Needless to say, all these
ingredients should be crucial for the possible supersymmetrization of
some $\tp{G}$-theories, most notably for the M-theory algebra $\tp{E_8}$. 

\subsection{Brane solutions}

There have been various attempts at relating $p$-brane solutions to
$\tp{G}$-theories \cite{Englert:2003py, West:2004iz, West:2004kb,
  Englert:2004it, Cook:2004er, Cook:2005wj}. Typically one postulates (extremal) brane solutions for positive (real) roots $\alpha$ of $\tp{G}$. Of
course not every root can correspond to a genuine brane solution, for
example in Minkowski signature extremal branes should have the
time-like direction contained in their world-volume.

In the literature also $S$-branes are discussed
\cite{Cook:2005wj}. The real solutions corresponding to these cannot
be extremal and hence are excluded from the category of solutions we
wish to discuss. There exist so-called extremal $S$-branes that are
essentially analytic continuations of the former real solutions, with
imaginary charges. For this we would have to allow for complex fluxes; this would lead to a (partial) analytic continuation of the algebra, and therefore to a relaxation of the strict conditions we have required for the representation of Poincar\'e duality and the real form of the $K(\tp{G})$-algebra's.  

The restrictions on the world volume signature of extremal branes found in
\cite{Hull:1998fh} can be neatly encoded in a simple constraint. It is
easily seen that, to a positive root there can be associated a genuine
extremal brane solution if and only if
\be \label{brane}
\inp{\alpha}{f} \ \mod \ 2= 1.
\ee
In essence this equation guarantees the right number of minus-signs in
the equation of motion. This simple equation covers both the minus
signs coming from the space-time signature, and possible explicit
signs in the kinetic terms in the Lagrangian. Because it is formulated
in terms of the invariant bilinear form, it transforms appropriately under Weyl
group transformations and outer automorphisms (and hence under T- and
S-dualities): If the root $\alpha$ gives a brane solution to a theory
with generalized space-time signature encoded in $f$, then $w(\alpha)$
gives a brane solution to a theory with space-time signature encoded
in $w(f)$, where $w$ is an element of the Weyl group or an outer
automorphism.

It is tempting to interpret $\inp{\alpha}{f}\  \mod \ 2=0$ as a condition for "instantonic", or S-branes. Strictly speaking however, the standard ansatz, with $i_0(f)=1$ and $\inp{\alpha}{f} \ \mod \ 2 =0$ does not lead to a solution. It was observed in
\cite{Cook:2005wj} that $S$-brane solutions can be interpreted as
solutions of theories in different space-time signatures. In the
present language this amounts to finding, for a given root $\alpha$,
an element $f$ of the coweight lattice that obeys $i_0(f) =1$ and
$\inp{\alpha}{f} \mod \ 2 =1$ simultaneously. There are in general
many solutions to these equations, defining just as many "Wick
rotations" of the original theory. In view of this non-uniqueness of
the Wick rotation, it is not quite clear what this continuation
exactly means.  

\subsection{The stability of Kaluza-Klein-theories}

It has been argued by Witten \cite{Witten:1981gj} that Kaluza-Klein
space-times are unstable. The original example studies the theory
arising from compactifying 5 dimensional General Relativity on a
circle. It is argued that this space-time decays by the nucleation of
"bubbles of nothing", with a decay time proportional to the radius of
the circle.

The "exotic dualities" of \cite{Englert:2003zs} and the suggestive
analysis of our section \ref{dual} seem to suggest that this theory
may allow a dual description, essentially as General relativity on a circle
of inverse radius. There is clearly some tension between this
T-duality, and the stability analysis: It seems impossible to
reconcile a decay time proportional to the radius of the circle, with
a dual description which essentially places the theory on a circle
with inverse radius. Of course we should not jump to such
conclusions. The analysis of \cite{Witten:1981gj} is only
semi-classical and valid in the limit of large radius, and should not
be na\"\i veley extrapolated to small radius. The arguments of section
\ref{dual} follow from a reckless extrapolation of semi-classical
results, which also has questionable aspects. 

Note that in the hypothetical T-duality, in the limit of small circle
size the Kaluza-Klein monopoles play a role similar to that of D0
branes in IIA-theory. Their charges organize to give the spectrum of a
Kaluza-Klein theory compactified on a dual circle, that decompactifies
when the radius of the original circle goes to zero. We cannot rely on
supersymmetry to claim that we have actually any control in the regime
of small circle size; this qualitative argument suggests however that
the original space-time may no longer be the appropriate reference
structure for such small circle sizes, because of the possible
existence of a dual space-time which is non-geometrical in the
original variables. It is clear that this would lead to novel
non-perturbative phenomena that have played no role in the analysis of
the paper \cite{Witten:1981gj}, and may offer new perspectives on
avoiding its conclusions.

It is furthermore amusing to note that the paper \cite{deBuyl:2005it}
and the present paper seem to suggest that the two description used in
\cite{Witten:1981gj}, the 5-dimensional Euclidean gravity on a circle
and the 5-dimensional Minkowski gravity on a circle are not only
analytic continuations of each other, but actually dual space-times
related by a T-duality. 

\subsection{Ghosts in Euclidean theories}

One of the main motivations for turning to theories in Euclidean
signature has always been that the path-integral, and various
quantities are better behaved in such signatures. In terms of
non-linear realizations this can be traced back to the compactness of
the Lorentz-group in these cases. 

For $\tp{G}$ algebra's, we have however argued that compact
denominator sub-algebra's are inconsistent with Poincar\'e duality. We
therefore either have to sacrifice the possibility of compact
$K(\tp{G})$, or the standard implementation of Poincar\'e
duality. There seems to be a sense in which minus-signs just cannot be
avoided in a hypothetical formulation of the theory with
$\tp{G}/K(\tp{G})$ symmetries. This raises interesting questions on
the existence of no-ghost theorems and the elimination of
non-physical degrees of freedom. Lacking a detailed formalism, it seems far too premature to discuss
such issues, but it is interesting to note that, in the context of
$\tp{G}$ algebra's we eventually have to face these sign-problems,
as dealing with them by analytic continuation leads at best to ambiguities.  

\subsection{Conclusions}

Theories that exhibit enhanced, hidden symmetries upon reduction,
allow oxidations to different higher dimensional theories. The least
one can say about $\tp{G}$ algebra's is that they form a bookkeeping
device that correctly records all the possible reductions and
oxidations. To correctly reproduce the space-time signatures and
explicit signs for these theories requires that Poincar\'e duality is
implemented correctly on these algebra's. We have argued in the
present paper that an element $f$ of the coweight lattice $Q(\tp{G})$
correctly encodes a space-time signature and sign-pattern consistent
with Poincar\'e duality, if and only if it satisfies the condition
\be \label{eyenaught}
i_0(f) =1.
\ee
Perhaps we should stress that this is \emph{not} a conjecture: In
spite of the fact that $\tp{G}$ algebra's are far from well
understood, the analysis we presented in section \ref{Gspecific},
reproduces all the results of section \ref{dual} which involved only
established techniques.

Of course the far more ambitious conjecture, that there should exist a
formulation of theories with (non-linearly realized, or linearized) $\tp{G}$ symmetries, would imply that the
ambiguities in reduction and oxidation correspond to physically
equivalent formulations of the theories. All these formulations would
then be related by "exotic T- and S-dualities". Even theories of
General Relativity have the appropriate structure to accommodate such
exotic symmetries, and they are suggestive of interesting
non-perturbative dynamics. Better still, a hypothetical
non-perturbative extension of such a theory has desirable properties,
namely the ability to avoid certain classes of singularities
(vanishing circles/cycles). This property is usually advertised as a
"good" property of string theories; trying to elevate it to a physical
principle would place the theories with enhanced symmetries, that
inspired triple-extended algebra's, in a privileged position. It would
be most interesting to check these hypothetical dualities in more
general situations, but it seems that the technology to do so
has not been yet developed.
 
Studying infinite-dimensional $\tp{G}$-algebra's and the corresponding
theories by level expansions gives a direct and intuitive link to
physics, but is necessarily limited to a finite number of levels, even
with powerful computer algorithms. Information that can be encoded in
an abstract fashion, such as (\ref{eyenaught}) that restricts the
signs to the ones consistent with Poincar\'e dualities, and the
condition for the existence of extremal branes (\ref{brane}), does not
refer to a specific theory or decomposition. We hope that more such
abstract, duality invariant tools will be developed; these should
provide us with insights that reach beyond any expansion to finite
level.

{\bf Acknowledgements:} 
I would like to thank Sophie de Buyl, Laurent Houart and Nassiba Tabti, for sharing and discussing the results of \cite{deBuyl:2005it} with me before publication, Axel Kleinschmidt for discussions, and Paul Cook and Peter West for correspondence.

The author is a post-doctoral researcher for the "FWO-Vlaanderen". This work was supported in part by the ``FWO-Vlaanderen'' through 
project G.0034.02, in part by the Belgian Federal Science Policy Office 
through the Interuniversity Attraction Pole P5/27 and in part by the 
European Commission FP6 RTN programme MRTN-CT-2004-005104.


\begin{thebibliography}{99}
\bibitem{T}
  J.~Polchinski,
  ``String theory. Vol. 1: An introduction to the bosonic string'';
  ``String theory. Vol. 2: Superstring theory and beyond'', Cambridge University Press, 1998;
  A.~Giveon, M.~Porrati and E.~Rabinovici,
  ``Target space duality in string theory,''
  Phys.\ Rept.\  {\bf 244}, 77 (1994)
  [arXiv:hep-th/9401139].

\bibitem{Cremmer:1979up}
E.~Cremmer and B.~Julia,
``The N=8 Supergravity Theory. 1. The Lagrangian,'' Phys.\ Lett.\ B
{\bf 80} (1978) 48; 
E.~Cremmer and B.~Julia, ``The SO(8) Supergravity,'' Nucl.\ Phys.\ B
{\bf 159} (1979) 141. 

\bibitem{Julia:1980gr}
B.~Julia, ``Group Disintegrations,''
in {\it C80-06-22.1.2} LPTENS 80/16
{\it Invited paper presented at Nuffield Gravity Workshop, Cambridge,
  Eng., Jun 22 - Jul 12, 1980}.

\bibitem{Julia:1982gx}
B.~Julia,
``Kac-Moody Symmetry Of Gravitation And Supergravity Theories,'' LPTENS 82/22
 {\it Invited talk given at AMS-SIAM Summer Seminar on Applications of
   Group Theory in Physics and Mathematical Physics, Chicago, Ill.,
   Jul 6-16, 1982}. 

\bibitem{Breitenlohner:1987dg}
P.~Breitenlohner, D.~Maison and G.~W.~Gibbons, ``Four-Dimensional
Black Holes From Kaluza-Klein Theories,'' Commun.\ Math.\ Phys.\  {\bf
  120} (1988) 295. 

\bibitem{Cremmer:1999du}
E.~Cremmer, B.~Julia, H.~Lu and C.~N.~Pope, ``Higher-dimensional
origin of D = 3 coset symmetries,'' arXiv:hep-th/9909099. 

\bibitem{Keurentjes:2002xc}
A.~Keurentjes,
``The group theory of oxidation,''
Nucl.\ Phys.\ B {\bf 658} (2003) 303
[arXiv:hep-th/0210178];

\bibitem{Keurentjes:2002rc}
  A.~Keurentjes,
  ``The group theory of oxidation. II: Cosets of non-split groups,''
  Nucl.\ Phys.\ B {\bf 658}, 348 (2003)
  [arXiv:hep-th/0212024].

\bibitem{Ortin:2004ms}
  T.~Ortin, ``Gravity and strings'', Cambridge University Press, 2004

\bibitem{Hull:1998vg}
C.~M.~Hull,
``Timelike T-duality, de Sitter space, large N gauge theories and  topological
field theory,''
JHEP {\bf 9807} (1998) 021
[arXiv:hep-th/9806146].

\bibitem{Hull:1998ym}
C.~M.~Hull,
``Duality and the signature of space-time,''
JHEP {\bf 9811} (1998) 017
[arXiv:hep-th/9807127].

\bibitem{Hull:1998fh}
C.~M.~Hull and R.~R.~Khuri,
``Branes, times and dualities,''
Nucl.\ Phys.\ B {\bf 536} (1998) 219
[arXiv:hep-th/9808069].

\bibitem{Cremmer:1998em}
  E.~Cremmer, I.~V.~Lavrinenko, H.~Lu, C.~N.~Pope, K.~S.~Stelle and T.~A.~Tran,
  ``Euclidean-signature supergravities, dualities and instantons,''
  Nucl.\ Phys.\ B {\bf 534}, 40 (1998)
  [arXiv:hep-th/9803259].

\bibitem{Hull:1998br}
C.~M.~Hull and B.~Julia,
``Duality and moduli spaces for time-like reductions,''
Nucl.\ Phys.\ B {\bf 534} (1998) 250
[arXiv:hep-th/9803239].

\bibitem{Keurentjes:2004xx}
  A.~Keurentjes,
  ``Time-like T-duality algebra,''
  JHEP {\bf 0411}, 034 (2004)
  [arXiv:hep-th/0404174].

\bibitem{West:2001as}
P.~C.~West,
``E(11) and M theory,''
Class.\ Quant.\ Grav.\  {\bf 18} (2001) 4443
[arXiv:hep-th/0104081];

\bibitem{Gaberdiel:2002db}
  M.~R.~Gaberdiel, D.~I.~Olive and P.~C.~West,
  ``A class of Lorentzian Kac-Moody algebras,''
  Nucl.\ Phys.\ B {\bf 645}, 403 (2002)
  [arXiv:hep-th/0205068].

\bibitem{West:2002jj}
  P.~West,
  ``Very extended E(8) and A(8) at low levels, gravity and supergravity,''
  Class.\ Quant.\ Grav.\  {\bf 20}, 2393 (2003)
  [arXiv:hep-th/0212291].

\bibitem{Englert:2003zs}
  F.~Englert, L.~Houart, A.~Taormina and P.~West,
  ``The symmetry of M-theories,''
  JHEP {\bf 0309}, 020 (2003)
  [arXiv:hep-th/0304206].

\bibitem{Kleinschmidt:2003mf}
  A.~Kleinschmidt, I.~Schnakenburg and P.~West,
  ``Very-extended Kac-Moody algebras and their interpretation at low  levels,''
  Class.\ Quant.\ Grav.\  {\bf 21}, 2493 (2004)
  [arXiv:hep-th/0309198].

\bibitem{Kac:1990gs}
  V.~G.~Kac,
  ``Infinite Dimensional Lie Algebras,'', Cambridge University Press, 1990.

\bibitem{West:2003fc}
P.~West,
``E(11), SL(32) and central charges,''
arXiv:hep-th/0307098.

\bibitem{Englert:2003py}
  F.~Englert and L.~Houart,
  ``G+++ invariant formulation of gravity and M-theories: Exact BPS
  solutions,''
  JHEP {\bf 0401}, 002 (2004)
  [arXiv:hep-th/0311255].

\bibitem{Schnakenburg:2003qw}
  I.~Schnakenburg and A.~Miemiec,
  ``E(11) and spheric vacuum solutions of eleven and ten dimensional
  supergravity theories,''
  JHEP {\bf 0405}, 003 (2004)
  [arXiv:hep-th/0312096].

\bibitem{Keurentjes:2003hc} 
A.~Keurentjes, 
``U-duality (sub-)groups and their topology,'' Class.\ Quant.\ Grav.\ {\bf 21}, S1367 (2004) 
[arXiv:hep-th/0312134].

\bibitem{Keurentjes:2004bv}
  A.~Keurentjes,
  ``E(11): Sign of the times,''
  Nucl.\ Phys.\ B {\bf 697}, 302 (2004)
  [arXiv:hep-th/0402090].

\bibitem{Englert:2004ph}
  F.~Englert, M.~Henneaux and L.~Houart,
  ``From very-extended to overextended gravity and M-theories,''
  JHEP {\bf 0502}, 070 (2005)
  [arXiv:hep-th/0412184].

\bibitem{deBuyl:2005it}
  S.~de Buyl, L.~Houart and N.~Tabti,
  ``Dualities and signatures of G++ invariant theories,''
  JHEP {\bf 0506}, 084 (2005)
  [arXiv:hep-th/0505199].

\bibitem{Cook:2005wj}
  P.~P.~Cook and P.~C.~West,
  ``M-theory solutions in multiple signatures from E(11),''
  arXiv:hep-th/0506122.

\bibitem{Barrett:1993yn}
  J.~W.~Barrett, G.~W.~Gibbons, M.~J.~Perry, C.~N.~Pope and P.~Ruback,
  ``Kleinian geometry and the N=2 superstring,''
  Int.\ J.\ Mod.\ Phys.\ A {\bf 9} (1994) 1457
  [arXiv:hep-th/9302073].

\bibitem{Helgason}
S.~Helgason, ``Differential geometry, Lie groups and symmetric
spaces,'', {\it New York, Academic Press (1978) (Pure and applied
  mathematics, 80)}.

\bibitem{Hinv}
  B.~de Wit and H.~Nicolai,
  ``D = 11 Supergravity With Local SU(8) Invariance,''
  Nucl.\ Phys.\ B {\bf 274} (1986) 363.
  H.~Nicolai,
  ``D = 11 Supergravity With Local SO(16) Invariance,''
  Phys.\ Lett.\ B {\bf 187} (1987) 316;
  B.~Drabant, M.~Tox and H.~Nicolai,
  ``Yet More Versions Of D = 11 Supergravity,''
  Class.\ Quant.\ Grav.\  {\bf 6} (1989) 255.

\bibitem{Damour:2002cu}
  T.~Damour, M.~Henneaux and H.~Nicolai,
  ``E(10) and a 'small tension expansion' of M theory,''
  Phys.\ Rev.\ Lett.\  {\bf 89}, 221601 (2002)
  [arXiv:hep-th/0207267].

\bibitem{Nicolai:2003fw}
  H.~Nicolai and T.~Fischbacher,
  ``Low level representations for E(10) and E(11),''
  arXiv:hep-th/0301017.

\bibitem{Gibbons:1979xm} 
G.~W.~Gibbons and S.~W.~Hawking, 
``Classification Of Gravitational Instanton Symmetries,'' Commun.\ Math.\ Phys.\ {\bf 66}, 291 (1979). 

\bibitem{Moore:1993zc}
  G.~W.~Moore,
  ``Finite in all directions,''
  arXiv:hep-th/9305139.


\cite{Vaula:2002cn}
\bibitem{Vaula:2002cn}
  S.~Vaula,
  ``On the construction of variant supergravities in D = 11, D = 10,''
  JHEP {\bf 0211} (2002) 024
  [arXiv:hep-th/0207080].

\bibitem{deBuyl:2005zy}
  S.~de Buyl, M.~Henneaux and L.~Paulot,
  ``Hidden symmetries and Dirac fermions,''
  Class.\ Quant.\ Grav.\  {\bf 22}, 3595 (2005)
  [arXiv:hep-th/0506009].


\bibitem{Nicolai:2004nv}
  H.~Nicolai and H.~Samtleben,
  ``On K(E(9)),''
  Q.\ J.\ Pure Appl.\ Math.\  {\bf 1}, 180 (2005)
  [arXiv:hep-th/0407055].

\bibitem{Keurentjes:2003yu}
  A.~Keurentjes,
  ``The topology of U-duality (sub-)groups,''
  Class.\ Quant.\ Grav.\  {\bf 21}, 1695 (2004)
  [arXiv:hep-th/0309106].


\bibitem{Miemiec:2004iv}
  A.~Miemiec and I.~Schnakenburg,
  ``Killing spinor equations from nonlinear realisations,''
  Nucl.\ Phys.\ B {\bf 698}, 517 (2004)
  [arXiv:hep-th/0404191].

\bibitem{West:2004iz}
  P.~West,
  ``Brane dynamics, central charges and E(11),''
  arXiv:hep-th/0412336.

\bibitem{West:2004kb}
  P.~West,
  ``E(11) origin of brane charges and U-duality multiplets,''
  JHEP {\bf 0408}, 052 (2004)
  [arXiv:hep-th/0406150].

\bibitem{Cook:2004er}
  P.~P.~Cook and P.~C.~West,
  ``G+++ and brane solutions,''
  Nucl.\ Phys.\ B {\bf 705}, 111 (2005)
  [arXiv:hep-th/0405149].

\bibitem{Englert:2004it}
  F.~Englert and L.~Houart,
  ``G+++ invariant formulation of gravity and M-theories: Exact  intersecting
  brane solutions,''
  JHEP {\bf 0405}, 059 (2004)
  [arXiv:hep-th/0405082].

\bibitem{Witten:1981gj} 
E.~Witten, 
``Instability Of The Kaluza-Klein Vacuum,'' 
Nucl.\ Phys.\ B {\bf 195} (1982) 481. 

\end{thebibliography}
\end{document}